\documentclass[a4paper,11pt]{article}
\pdfoutput=1 

\usepackage{jcappub} 

\usepackage[T1]{fontenc} 
\usepackage{caption}
\usepackage{subcaption}
\usepackage[section]{placeins}
\usepackage{float}

\title{\boldmath Constraining Ultralight Axions with Galaxy Surveys}

\author[a,b,c,1]{A. Lagu\"e,\note{Corresponding author.}}
\author[b]{J. R. Bond,}
\author[a,c]{R. Hlo\v zek,}
\author[c]{K. K. Rogers,}
\author[d]{D. J. E. Marsh,}
\author[e]{and D. Grin}

\affiliation[a]{David A. Dunlap Department of Astronomy \& Astrophysics, University of Toronto, 50 St. George St., Toronto, ON, M5S 3H4, Canada}
\affiliation[b]{Canadian Institute for Theoretical Astrophysics, University of Toronto, 60 St. George St., Toronto, ON, M5S 3H8, Canada}
\affiliation[c]{Dunlap Institute for Astronomy and Astrophysics, University of Toronto, 50 St. George St., Toronto, ON, M5S 3H4, Canada}
\affiliation[d]{Institut f\"ur Astrophysik, Georg-August Universit\"at, Friedrich-Hund-Platz 1, G\"ottingen, Germany}
\affiliation[e]{Department of Physics and Astronomy, Haverford College, Haverford, PA 19041, USA}

\emailAdd{lague@cita.utoronto.ca}

\abstract{Ultralight axions and other bosons are dark matter candidates present in many high energy physics theories beyond the Standard Model. In particular, the string axiverse postulates the existence of up to $\mathcal{O}(100)$ light scalar bosons constituting the dark sector. Considering a mixture of axions and cold dark matter, we obtain upper bounds for the axion relic density $\Omega_a h^2 < 0.004$ for axions of mass $10^{-31}\;\mathrm{eV}\leq m_a \leq 10^{-26}\;\mathrm{eV}$ at 95\% confidence. We also improve existing constraints by a factor of over 4.5 and 2.1 for axion masses of $10^{-25}$ eV and $10^{-32}$ eV, respectively. We use the Fourier-space galaxy clustering statistics from the Baryon Oscillation Spectroscopic Survey (BOSS) and demonstrate how galaxy surveys break important degeneracies in the axion parameter space compared to the cosmic microwave background (CMB). We test the validity of the effective field theory of large-scale structure approach to mixed ultralight axion dark matter by making our own mock galaxy catalogs and find an anisotropic ultralight axion signature in the galaxy quadrupole. We also observe an enhancement of the linear galaxy bias from 1.8 to 2.4 when allowing for 5\% of the dark matter to be composed of a $10^{-28}$ eV axion in our simulations. Finally, we develop an augmented interpolation scheme allowing a fast computation of the axion contribution to the linear matter power spectrum leading to a 70\% reduction of the computational cost for the full Monte Carlo Markov chains analysis.}

\begin{document}
\maketitle
\flushbottom

\section{Introduction}
\label{sec:intro}
Multiple hypotheses have been put forward to explain the nature of dark matter. Among the most promising candidates from particle physics are axions which were initially theorized as a solution to the charge parity problem in quantum chromodynamics (QCD). A cosmological population of such axions produced by vacuum realignment~\cite{Abbott1983ACosmological,Preskill1983Cosmologyof,Dine1983TheNot} would have a relic density given by~\cite{Ballesteros2017StandardModel}
\begin{align}
    \Omega_{a}^{\mathrm{QCD}} h^2 \approx 0.12\left( \frac{6\;\mu\mathrm{eV}}{m_{a}}\right)^{1.165} \theta_I^2,
\end{align}
where $\theta_I$ is the initial misalignment angle and $h$ is the Hubble constant today in units of 100 km/s/Mpc, and $m_a$ is the axion mass. We note that the QCD axion has a lower mass bound of around $m_a\sim 6\;\mu$eV for $\theta_I=\mathcal{O}(1)$ to avoid the relic density from exceeding the cosmological dark matter density $\Omega_dh^2\approx 0.12$. Escapes from this can be found in the so-called ``anthropic'' axion window by fine-tuning the initial axion field. Indeed, assuming the PQ symmetry is broken after inflation, it is possible to prevent the lighter QCD axion from exceeding the total dark matter relic density by appropriately reducing the initial misalignment angle~\cite{Wilczek2004AModel}.

On the other hand, axions also appear naturally in string theory where they can be many orders of magnitude lighter than their QCD counterparts leading to the concept of an axiverse~\cite{Arvanitaki2010StringAxiverse}. The relic density in this regime follows (assuming again a vacuum realignment formation mechanism)~\cite{Tanabashi2018ReviewOf}
\begin{align}
    \Omega_{a} h^2 = 0.12 \left(\frac{m_{a}}{4.7\times 10^{-19}\;\mathrm{eV}}\right)^{1/2} \left(\frac{f_a}{10^{16}\;\mathrm{GeV}}\right)^2\left( \frac{\Omega_mh^2}{0.15}\right)^{3/4} \left(\frac{3.4\times 10^3}{1+z_\mathrm{eq}}\right)^{3/4}\theta_I^2, \label{eq:ax_dens}
\end{align}
where $f_a$ is the symmetry breaking scale of the axion, $\Omega_m$ is the total matter density, and $z_\mathrm{eq}$ is the redshift of matter-radiation equality. In the latter case, they are theorized to form in a \textit{plenitude} of $\mathcal{O}(100)$ logarithmically distributed masses~\cite{Arvanitaki2010StringAxiverse,Bachlechner2017MultipleAxion}. Ref.~\cite{Mehta2021SuperradianceIn} studied the axion mass and decay constant distributions in Type IIB compactifications on Calabi-Yau manifolds and found and almost log-flat distributions for $m_a$, and log-normal distributions for $f_a$. The implications of a single ultralight scalar boson of mass $m_a\sim 10^{-22}$ eV have been explored in the context of fuzzy dark matter (FDM) (see e.g. Ref.~\cite{Hui2021WaveDark} for a review). This scenario assumes however that a single light boson composes exactly all of the dark matter. Given the form of Eq.~(\ref{eq:ax_dens}), we can ask what happens if we relax this assumption and allow multiple axions each composing a fraction of the dark matter. Many studies have been devoted to probing the existence of such particles as a subdominant component~\cite{Grin2019GravitationalProbes,Diehl2021ConstrainingUltra-light} and any evidence for such a multi-component dark sector would be considered a smoking gun for the string landscape~\cite{Arvanitaki2010StringAxiverse}. 

There is also motivation for the existence of light bosonic dark matter from a cosmological perspective as tensions arise between the predictions of cold dark matter (CDM)-only simulations and observations. This is known as the small-scale crisis of CDM. Notably, there is the missing satellite problem where the number of satellite galaxies predicted by CDM is much larger than what has been historically observed~\cite{Klypin1999WhereAre,Moore1999DarkMatter}. There is also the cusp-core problem where halos in CDM simulations have a diverging density profile at their center when observations seem to suggest  the existence of central cores with an approximately constant density~\cite{deBlok2010TheCore}. Finally, there is the too-big-to-fail problem which describes the fact that some CDM subhalos are theoretically too massive not to have formed stars and should be observationally detectable~\cite{Boylan-Kolchin1999TooBig}. Many modifications to CDM have been proposed to alleviate these tensions such as self-interacting dark matter~\cite{Spergel2000ObservationalEvidence} and warm dark matter~\cite{Sommer-Larson2001FormationOf}. Also, detailed hydrodynamical simulations have shown that baryonic effects such as feedback and active galactic nuclei solve many of these issues~\cite{Pontzen2012MNRAS.421.3464P, Pontzen2014Natur.506..171P, Tollet2016NihaoIV,Fitts2017FireIn}. For ultralight axions, it has been shown that their lack of small-scale clustering and formation of solitonic cores could offer a solution to all of the above tensions~\cite{Marsh2014AModel,Marsh2015AxionDark}. It has been argued in previous work that the existence of ultralight bosons of mass $m_a\sim 10^{-22}$~eV could potentially exacerbate the cusp-core problem by creating overly dense cores~\cite{Robles2019ScalarField} and predict dwarf galaxies that are too massive~\cite{Safarzadeh2020Ultra-Light}. However, these conclusions are based on the assumption that the dark matter is composed of a single axion species whereas high energy physics naturally allows a number of axion species $N_\mathrm{ax}\gg 1$.

Existing bounds on ultralight axion dark matter (\(m_a \lesssim 10^{-20}\,\mathrm{eV}\)) exploit their distinctive effect on the cosmic large-scale structure. This is most prominently a suppression in the growth of structure, relative to the CDM limit, below a characteristic scale owing to so-called ``quantum pressure.'' This manifests as a cut-off in the linear matter power spectrum, where the cut-off wavenumber is a monotonically increasing function of the axion mass \citep{Hu2000FuzzyCold}. On the largest observable scales, the cosmic microwave background (CMB) rules out axions being all the dark matter for \(10^{-33}\,\mathrm{eV} \leq m_a \leq 10^{-24}\,\mathrm{eV}\) and places percent-level bounds on their fractional contribution to the dark matter \citep{Hlozek2018UsingMatter}.\footnote{The CMB is also sensitive to axion-induced isocurvature modes and the mass regime where ultralight axions behave like dark energy.} On the smallest scales currently accessible in the quasi-linear matter power spectrum \citep[\(k \sim 20\,h\,\mathrm{Mpc}^{-1}\);][]{Boera2019ApJ...872..101B, Chabanier2019MNRAS.489.2247C}, the Lyman-alpha forest \citep{Irsic2017PhRvL.119c1302I, Kobayashi2017LymanAlpha, Armengaud2017MNRAS.471.4606A} sets the strongest bound on the mass scale (\(m_a \sim 10^{-22}\,\mathrm{eV}\)) motivated above, allowing axions to be all the dark matter only if \(m_a > 2 \times 10^{-20}\,\mathrm{eV}\) \citep{Rogers2021PhRvL.126g1302R, Rogers2021PhRvD.103d3526R}. In setting this bound, it is important to marginalise carefully the partly degenerate suppression scales arising from the temperature and pressure of the intergalactic medium \citep{Lukic2015MNRAS.446.3697L, Onorbe2017ApJ...837..106O, Rogers2021PhRvD.103d3526R}. Competitive bounds also arise from not observing an axion-induced suppression in the Milky Way sub-halo mass function (\(m_a > 2.9 \times 10^{-21}\,\mathrm{eV}\)), with a different set of nuisance parameters and systematic effects \citep{Nadler2021PhRvL.126i1101N}. Also note that the Lyman-$\alpha$ forest bounds the axion fraction in the mixed dark matter scenarios, but only at the $\approx30$\% level~\cite{Kobayashi2017LymanAlpha}. X-ray observations have also been used to make constraints on ultralight axions as dark matter reaching a constraint of $m_a>7\times 10^{-23}$ eV~\cite{Maleki2020InvestigationOf}. In this work, we exploit a different regime in the matter power spectrum, as well as, for the first time, the anisotropic effect on matter clustering.

\begin{figure}
    \centering
    \includegraphics[width=0.85\linewidth]{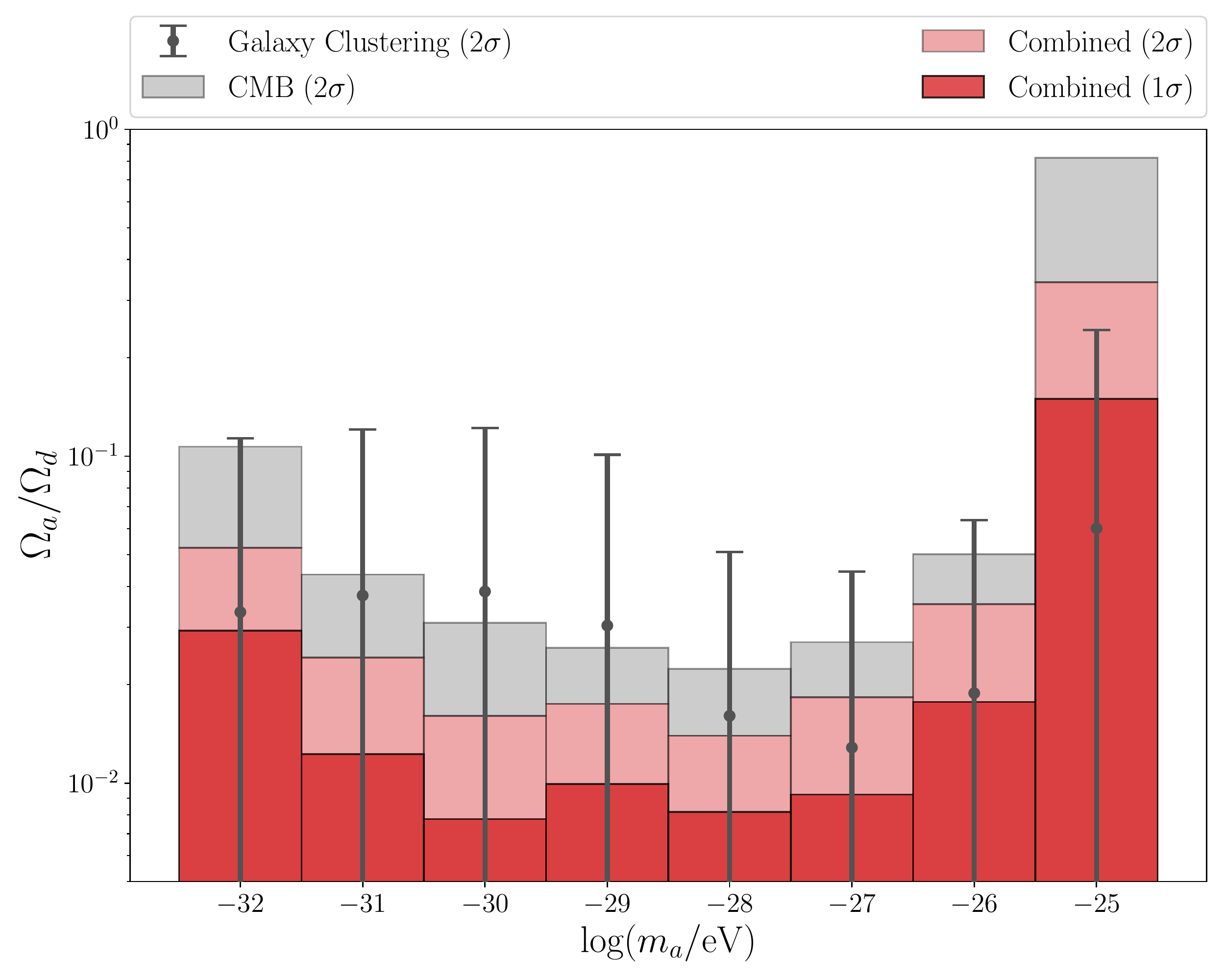}
    \caption{Constrained region of the $m_a-\Omega_a/\Omega_d$ parameter space when combining galaxy clustering and CMB data. The light and dark red regions represent the $2\sigma$ and $1\sigma$ allowed regions respectively. The grey shaded area denotes the boundary of the $2\sigma$ allowed region using the Planck 2015 data release from the analysis of Ref.~\cite{Hlozek2015AData} and the grey error bars denote the 2$\sigma$ bound found using only galaxy surveys as part of the present work. Previous work combining the CMB and galaxy clustering data found $\Omega_a/\Omega_d \leq 0.048$ for a similar axion mass range~\cite{Hlozek2015AData} while we find that a $10^{-28}$ eV axion has $\Omega_a/\Omega_d \leq 0.014$ at 95\% confidence. The wider range of allowed $\Omega_a/\Omega_d$ when adding CMB prior at $10^{-25}$ eV stems from a degeneracy with the power spectrum amplitude (see Fig.~\ref{fig:1_25_degen}).}
    \label{fig:combined_constraint}
\end{figure}

Galaxies are tracers of the matter power spectrum and deviations from the CDM power spectrum due to the presence of ultralight axions has a characteristic signature which we can probe using galaxy clustering data. In order to constrain small concentrations of ultralight axions, we measure the maximal amount of suppression of the galaxy power spectrum from the BOSS DR12 survey for various axion scenarios. Without loss of generality, we consider only one ultralight axion species while the remainder of the dark matter is treated as pure CDM. This model is equivalent to a multi-axion scenario where only one of the axion masses falls in the ultralight regime. Deviations from CDM would be even larger in the presence of multiple ultralight fields (assuming the same relic density) which implies that the single axion + CDM model in this study will lead to conservative constraints. In Section~\ref{sec:ULA_phys}, we describe the physics of ultralight axions and their behaviour in more detail. Next, in Section~\ref{sec:model}, we present the galaxy clustering model used in this work while the full set of parameters and priors are described in Section~\ref{sec:methods}. In Section~\ref{sec:simtest}, we validate our model with large-scale structure simulations both with and without an axionic component. Finally, we consider the results of the full likelihood analysis in Section~\ref{sec:results}, and discuss the implications of our findings in Section~\ref{sec:discussion}. In Appendix~\ref{app:perturbations} we develop our model in greater detail and examine the impact of the wave effects on higher order corrections to the power spectrum, in Appendix~\ref{app:anisotropic} we study the impact of the power spectrum suppression on the galaxy quadrupole, and Appendix~\ref{app:axion_interpolation} we describe our optimized axion transfer function interpolation scheme.


\section{Ultralight Axion Physics}
\label{sec:ULA_phys}
Ultralight axions are scalar bosons described by a non-relativistic field $\phi$ with a cosine potential which we approximate as a harmonic quadratic potential of the form
\begin{align}
    V(\phi) &\propto 1- \cos(\phi/f_a)
    \\&\approx \frac{1}{2}m_a^2 \phi^2.
\end{align}
This assumes $\phi\ll f_a$ which is consistent with the treatment of Ref.~\cite{Hlozek2015AData,Hlozek2018UsingMatter}\footnote{See Ref.~\cite{Leong2019TestingExtreme} for certain cases where $\phi~\sim f_a$ could be advantageous.}. Across the mass range of interest, this approximation is valid for $f_a$ slightly above the Grand Unified Theory (GUT) scale (see e.g. Ref.~\cite{Bauer2021IntensityMapping}). The corresponding density and pressure of the background axion field are then given respectively by
\begin{align}
    \rho_a &= \frac{1}{2a^2}\dot{\phi_0}^2 + \frac{m_a^2}{2}\phi_0^2, \\
    P_a &= \frac{1}{2a^2}\dot{\phi_0}^2 - \frac{m_a^2}{2}\phi_0^2,
\end{align}
where the dot denotes the derivative with respect to conformal time and $\phi_0$ is the homogeneous background field. The equation of motion of the field for the field reads
\begin{align}
    \ddot{\phi_0} + 2\mathcal{H} \dot{\phi_0} + \frac{1}{2}a^2 m_a^2 \phi_0 =0,
\end{align}
where $\mathcal{H}=aH$ is the conformal Hubble factor. At early times ($H\gg m_a$), the field is slowly rolling with $\dot{\phi}\approx 0$, the axion equation of state is $w_a \equiv P_a/\rho_a\approx -1$, and the field behaves as a dark energy component. At late times ($H\ll m_a$), the field's equation of state oscillates around zero and the energy density of the field obeys the scaling $\rho_a\propto a^{-3}$. It is therefore a dark matter component after the field started oscillating around the potential minimum. The value of the scale factor when this transition happens is denoted $a_\mathrm{osc}$ and defined such that
\begin{align}
    3H(a_\mathrm{osc}) \approx m_a.
\end{align}
This implies that the time of transition of the field is given as a function of its mass. For $m_a\lesssim 10^{-28}$ eV, and $a_\mathrm{osc}>a_\mathrm{eq}$ the field does not behave as dark matter at matter-radiation equality and cannot constitute the entirety of the dark matter. We refer to those axions as dark-energy-like axions. For any mass, the mean axion density depends on $a_\mathrm{osc}$ and is given by
\begin{align}
    \rho_a = a_\mathrm{osc}^3\bigg[\frac{1}{2a^2}\dot{\phi_0}^2 + \frac{m_a^2}{2}\phi_0^2 \bigg]_{m_a=3H}.
\end{align}

One of the main signatures of axions is their lack of clustering on small scales. Axions have a characteristic Jeans scale $k_{J}$ below which the growth of structure is suppressed\footnote{See Ref.~\cite{Khlopov1985GravitationalInstability} for a study of the Jeans instability in the presence of self-interactions.}. For the linear matter power spectrum with dark-matter-like axions, the suppression is frozen in at matter-radiation equality and we have~\cite{Hu2000FuzzyCold}
\begin{align}
    k_{J,\mathrm{eq}} \approx 9 \bigg(\frac{m_a}{10^{-22}\;\mathrm{eV}}\bigg)^{1/2} \mathrm{Mpc}^{-1}.
\end{align}

\begin{figure}
    \centering
    \includegraphics[width=\linewidth]{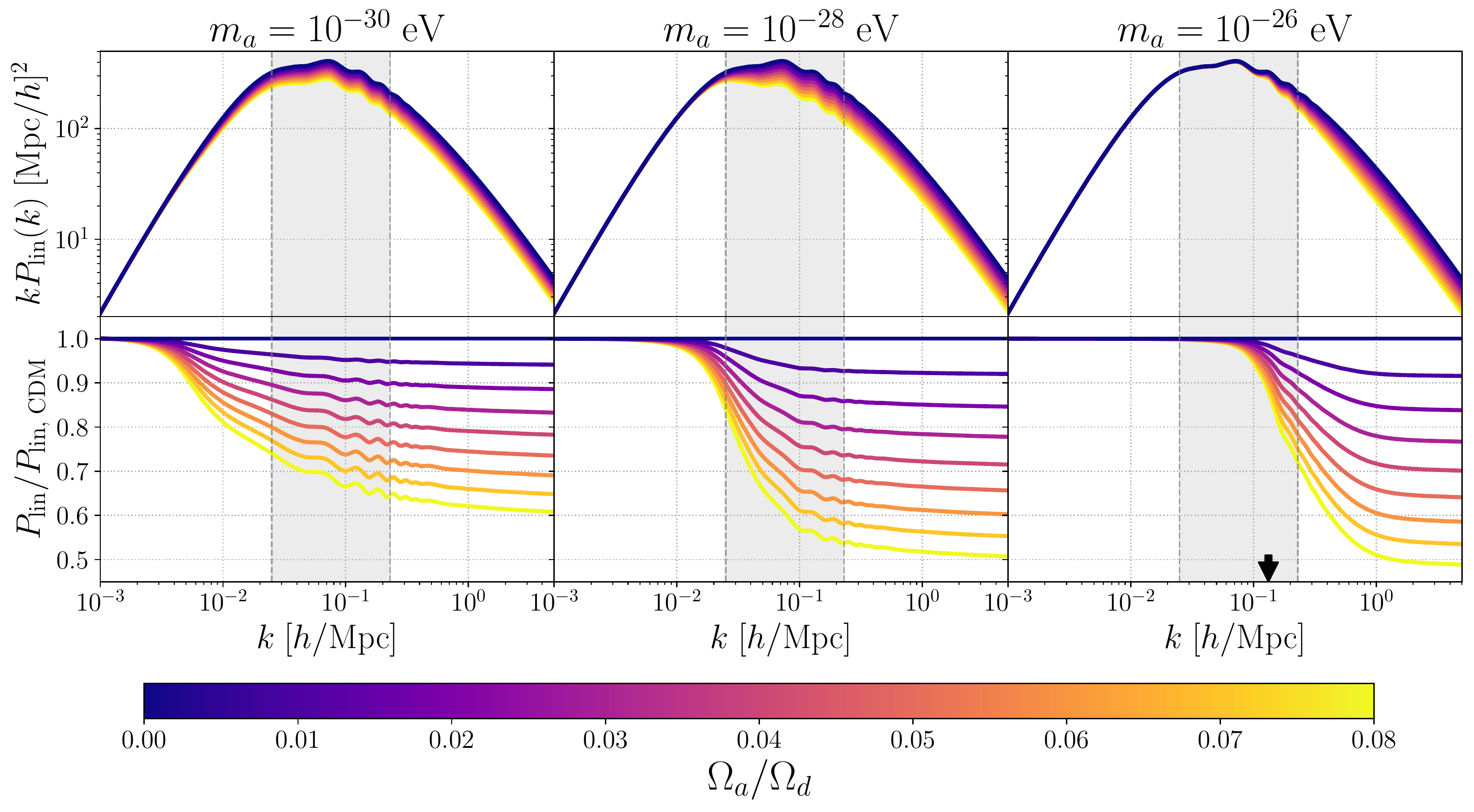}
    \caption{Linear power spectrum for different axion masses and densities. We note that the level of suppression increases with the axion density while the scale at which the axion power spectrum deviates from the CDM ($\Omega_a=0$) case is dependent on the axion mass. The grey shaded area highlights the Fourier modes probed by our analysis between $k_\mathrm{min}=0.025\;h/$Mpc and $k_\mathrm{max}=0.23\;h/$Mpc. The black arrow indicates the Jeans scale for the only axion mass for which $a_\mathrm{osc} < a_\mathrm{eq}$ among the three showed. For reference, the massive neutrino characteristic scale of $k_\mathrm{nr}$ for a neutrino with a mass of $150$ eV would correspond to the Jeans scale of axions of mass $10^{-26}$ eV.}
    \label{fig:linear_pk}
\end{figure}
This holds when assuming that the axions behave as dark matter at matter-radiation equality or that $m_a\gtrsim 3H(a_\mathrm{osc})$. In the alternate case, the field behaves as a dark energy component until it starts oscillating. The Jeans scale at matter-radiation equality is not the optimal reference scale for these axions. We need instead to use the scale at horizon crossing when the field started oscillating which we can calculate as~\cite{Amendola2006DarkMatter,Arvanitaki2010StringAxiverse,Bauer2021IntensityMapping}
\begin{align}
    k_m = a_\mathrm{osc} H\left(a_\mathrm{osc}\right).
\end{align}
We note that in both cases, the loss in power on small scales is dependent on the axion fraction $\Omega_a/\Omega_d$ where $\Omega_d$ is the total dark matter density (combining CDM and axions). The level of suppression may then only be partial when fixing the axion mass and varying the axion concentration with the $\Omega_a\to 0$ limit recovering the pure CDM result. This deviation from CDM in the linear power spectrum illustrated in Fig.~\ref{fig:linear_pk} is the main signature we use in this study to constrain axions. The matter power spectra including axionic effects are computed using the adapted Boltzmann code \texttt{axionCAMB}~\cite{Hlozek2015AData}. It is worth noting that the code solves for the evolution of the axion perturbations using an effective fluid approach. It has been shown to match closely the exact solution, although biases could occur if the sensitivity of the experiments used approaches the cosmic variance limit~\cite{Cookmeyer2020HowSound}.

\section{Model}
\label{sec:model}
Our datasets consist of the multipole moments of the CMASS ($z_\mathrm{eff}=0.57$) and LOWZ ($z_\mathrm{eff}=0.32$) catalogs from the BOSS survey~\cite{Beutler2017TheClustering}. We use the North and South Galactic Caps denoted NGC and SGC for our analysis. Given galaxy positions, the local galaxy overdensity is defined as
\begin{align}
    \delta_g(\mathbf{x}) \equiv \frac{n_g(\mathbf{x})}{\bar{n}_g}-1,
\end{align}
where $n_g(\mathbf{x})$ is the number density of galaxies at the comoving position $\mathbf{x}$ and $\bar{n}_g$ is the mean number density for the survey volume $V$. After taking the Fourier transform $\delta_g(\mathbf{k})$ of the real-space overdensity, the multipole moments $P_\ell$ of the power spectrum are calculated by integrating over the angle between the line-of-sight and the wavevector $\mathbf{k}$. It is customary to decompose the latter into $k$ which is the scalar amplitude of the vector and $\mu$ which is the cosine of angle between the wavevector and the line-of-sight. The integral expression for the galaxy multipoles can then be written as (using the Yamamoto estimator~\cite{Yamamoto2006AMeasurement})
\begin{align}
    P_\ell (k) = \left\langle \frac{2\ell+1}{V}\int\frac{d\Omega_k}{4\pi} \delta_g(\mathbf{k})\delta_g(-\mathbf{k}) \mathcal{P}_\ell\left(\hat{\mathbf{k}}\cdot\hat{\mathbf{z}}\right)\right\rangle,
    \label{eq:P_ell}
\end{align}
where the hat denotes unit vectors and $\mathcal{P}_\ell$ are the Legendre polynomials of degree $\ell$. In the data, the shot noise $P_\mathrm{shot} = 1/\bar{n}_g$ is subtracted from the monopole. For the purpose of this study, we will focus on the monopole ($\ell=0$) and quadrupole ($\ell=2$) moments of the power spectrum.

\begin{figure}
    \centering
    \includegraphics[width=\linewidth]{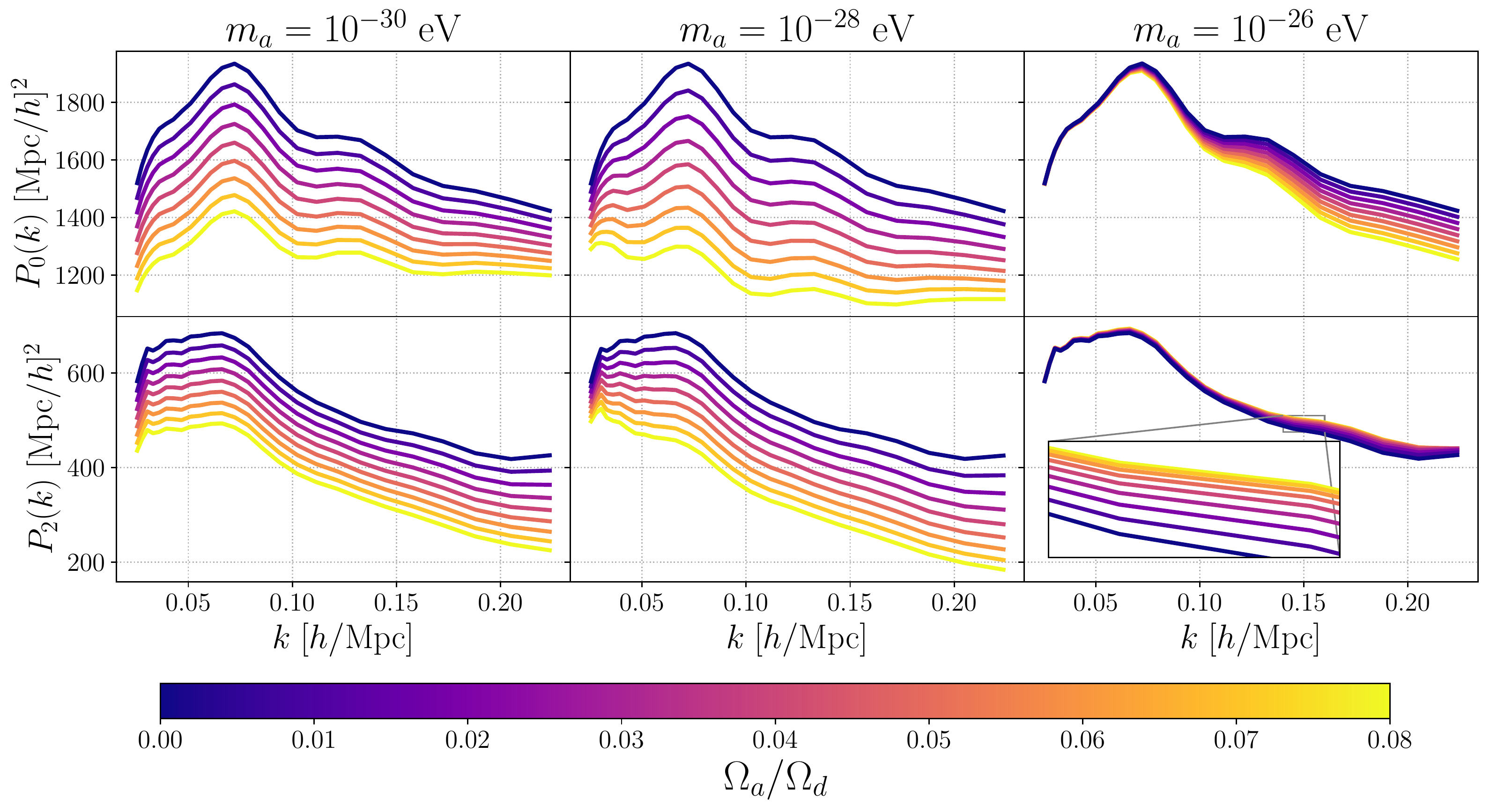}
    \caption{Monopoles and quadrupoles of the galaxy clustering obtained from the linear power spectra of Fig.~\ref{fig:linear_pk}. The model is evaluated with fixed cosmological and bias parameters at redshift $z=0.57$. We highlight a slight increase in the quadrupole on small scales indicating that the structure suppression from axions also has anisotropic effects.}
    \label{fig:mon_quad_model}
\end{figure}

Our model for the multipoles begins with the linear matter power spectrum generated from a set of standard flat $\Lambda$CDM cosmological parameters and a set of two axion parameters. We first obtain the power spectrum without the axion effects using the Boltzmann code \texttt{Class}~\cite{Blas2011TheCosmic} and the axion transfer function with the adapted code \texttt{axionCAMB}. We refer to this power spectrum as the CDM power spectrum. The reason for this use of code combination is to separate the calculations for the axion effects from the rest of the cosmological calculations since they take significantly longer to complete. We opt to interpolate over the axion transfer function rather than the full matter power spectrum since the axion transfer function is independent of  $A_s$. This implies that we can reduce the dimensionality of our interpolation tables for the axion transfer function by interpolating only over the axion fraction, the Hubble constant, the baryon density, and the dark matter density. This saves considerable computational resources. We obtain percent-level errors compared to the use of the full \texttt{axionCAMB} when the axion fraction is less than 10\%. As for the choice of Boltzmann code for the $\Lambda$CDM part of the calculation, we can use any code of preference since the CDM power spectrum has been factored out of the axion transfer function and since the latter is only weakly dependent on the choice of cosmology at low axion fractions (we explore this more rigorously in Appendix~\ref{app:axion_interpolation}). Ultimately, the choice of the \texttt{Class} code was made due to the fact that it is the one implemented with the galaxy multipole calculator which we will use for our analysis.

We then load the interpolation tables and use them to compute the axion transfer function so that the linear power spectrum with axions is given by
\begin{align}
    P_\mathrm{lin}(k,z) = T_\mathrm{ax}^2\left(k,z;m_a,\Omega_a/\Omega_d\right) P_\mathrm{lin,\;CDM}(k,z).\label{eq:transfer}
\end{align}
Examples of $P_\mathrm{lin}$ and $T_\mathrm{ax}^2$ are given in the top and bottom rows of Fig.~\ref{fig:linear_pk} respectively. The computation of the axion transfer function interpolation tables is given in more detail in Appendix~\ref{app:axion_interpolation}.

The redshift-space galaxy power spectrum can be summarized as a sum of corrections to the (biased) linear matter power spectrum (including axion effects)
\begin{align}
    P_g(\mathbf{k}) = b_1^2 P_\mathrm{lin}(k) + P_\mathrm{1-loop}(\mathbf{k})+P_\mathrm{counter}(\mathbf{k})+P_\mathrm{stoch}(k),
\end{align}
where the contributions are respectively from the 1-loop corrections, the counterterms, and the stochastic terms. This model is based on the Effective Field Theory of Large Scale Structure (EFTofLSS)~\cite{Carrasco2012TheEffective} which allows high accuracy prediction of the power spectrum up to semi-linear scales. This approach consists in perturbatively expanding the equations for the power spectrum corrections in powers of $k/k_\mathrm{NL}$ where $k_\mathrm{NL}$ is the scale at which non-linear effects begin to dominate. It accounts for physics on smaller scales through the counterterms while stochastic terms model the variations between the expectation values of the stress tensor and its value in a given realization. For more details on the EFTofLSS, we direct the readers to Ref.~\cite{Carrasco2012TheEffective} and to subsequent publications.

The EFTofLSS was designed with the assumption of collisionless dark matter. In the case of ultralight axions of mass $m_a \leq 10^{-25}$ eV, the dark matter particles do not cluster on small scales due to their scale-dependent sound speed~\cite{Chavanis2012GrowthMatter,Suarez2015HydrodynamicAnalysis,Marsh2016AxionCosmology}. For this reason, we may treat them like fast moving neutrinos even if the axions are strictly speaking non-relativistic. The inclusion of fast moving neutrinos in the EFTofLSS was performed in Ref.~\cite{Senatore2017TheEffective}. 
It was found that at first order, the supplementary counterterms that need to be added to account for neutrinos have the same functional form as the counterterms already present. Given that we vary the prefactors of these counterterms with a loose prior in our analysis, we assume that the functional form of our counterterms accounts for the non-clustering axions. Therefore the EFTofLSS can be extended to ultralight axions when marginalizing over the amplitude of the counterterms (which are defined in more detail in Appendix~\ref{app:perturbations}). Furthermore, we explore the impact of the wave effects on the clustering dynamics on higher order corrections in Appendix~\ref{app:perturbations} while we leave the computation of the counterterms including axion wave effects for future work. We conclude that the deviations from pure CDM dynamics are a small contribution to already subdominant higher order corrections, especially at low axion fractions. 

We then use the public \texttt{PyBird} code to apply the following transformations to the linear matter power spectrum (with the corresponding reference in the literature):
\begin{enumerate}
    \item compute and add the non-linear one-loop correction to the power spectrum~\cite{Perko2016BiasedTracers},
    \item perform the IR-resummation of the power spectrum~\cite{Ivanov2018InfraredResummation},
    \item apply window function effects~\cite{Zheng2019TheClustering},
    \item apply the Alcock-Paczynski effect~\cite{DAmico2020TheCosmological} ,
    \item account for power spectrum binning~\cite{DAmico2020TheCosmological},
    \item correct for fiber collisions~\cite{DAmico2020TheCosmological}.
\end{enumerate}
In the following section, we will briefly discuss each of the steps performed on the linear matter power spectrum to obtain a realistic model for the galaxy multipole data. The calculations for each of these terms from the parameters of Table~\ref{tab:priors} are explained in greater detail in Appendix~\ref{app:perturbations}. Examples of galaxy multipoles obtained with our model including axions with fixed nuisance parameters are shown in Fig.~\ref{fig:mon_quad_model}.

The \texttt{PyBird} code has the convenient feature of computing the effect of window functions from surveys. We use this when running the tests on the simulated data. However, the dataset we use from BOSS combines the North and South Galactic Caps for each redshift slice. We therefore use a custom window function computation. We obtain the monopole and quadrupole with the publicly available window matrices $W_{ij}$ using Ref.~\cite{GilMarin2015TheClustering} Eq.~(15)
\begin{align}
    P_{\ell}(k_i) = \sum_j W_{ij}^{\ell 0} P_{0}^\mathrm{no-window}(k_j) + \sum_j W_{ij}^{\ell 2} P_{2}^\mathrm{no-window}(k_j),
\end{align}
where $k_i$ are the $k$-bins of the BOSS dataset and $k_j$ are 1000 selected $k$-bins up to 0.5 $h$/Mpc.



\section{Methodology}\label{sec:methods}
Our model based on the EFTofLSS was described in Section~\ref{sec:model} and has 10 free parameters. The monopole and quadrupole are both calculated at redshifts $z=0.32$ and $z=0.57$. For each likelihood calculation, the cosmological parameters are assumed to be the same for the two redshift slices, but the bias, counterterms, and stochastic parameters are allowed to vary. The axion fraction is varied like the cosmological parameters while the axion mass is fixed for each individual calculation. We thus make sixteen independent likelihood calculations: for each of the eight mass bins between $10^{-32}$ eV and $10^{-25}$ eV, we compute the posterior distribution for two choices of priors. For the first half of the runs, we include only information about galaxy clustering and use a Big Bang Nucleosynthesis (BBN) prior on the baryon density $\omega_b$. For the second set of runs which combine our setup with existing constraints, we instead impose a CMB prior on all cosmological parameters along with a prior on the axion fraction also from the same CMB analysis. We thus use the pre-computed chains from the Planck likelihood including an axion component directly. This method has been used as an alternative to joint likelihood analysis when combining large-scale structure and CMB data with the EFTofLSS (see Ref.~\cite{Ivanov2020CosmologicalParameters,DAmico2020TheCosmological} where a CMB prior is put on the baryon density and sound horizon).


\begin{table}
    \centering
    \begin{tabular}{ |ccc| } 
        \hline
        Parameter & Type & Prior \\
        \hline
        $100\omega_b$ & Cosmology & $\mathcal{N}(2.214,\;0.038)$ \\
        $\omega_d$ & Cosmology & $\mathcal{U}(0.05,\;0.2)$ \\
        $\ln(10^{10}A_s)$ & Cosmology & $\mathcal{U}(\log(5),\;\log(50))$ \\
        $h$ & Cosmology & $\mathcal{U}(0.55,\;.91)$ \\
        \hline
        $m_a$ & Axion & Discrete Bins  \\
        $\Omega_a/\Omega_d$ & Axion & $\mathcal{U}(0,\;1)$  \\ 
        \hline
        $b_1$ & Bias & $\mathcal{U}(0,\;4)$   \\
        $b_2$ & Bias & $\mathcal{U}(-2\sqrt{2},\;2\sqrt{2})$   \\
        $b_3$ & Bias & $\mathcal{N}(0,\;2)$   \\
        $b_4$ & Bias & Fixed (equal to $b_2$)   \\
        \hline
        $c_\mathrm{ct}$ & Counterterm & $\mathcal{N}(0,\;2)$  \\
        $\tilde{c}_{r,1}$ & Counterterm & $\mathcal{N}(0,\;8)$  \\
        $\tilde{c}_{r,2}$ & Counterterm & Fixed to 0  \\
        \hline
        $\tilde{c}_{\epsilon, 1}/1000$ & Stochastic & $\mathcal{U}(-10,\;10)$  \\
        $\tilde{c}_{\epsilon ,2}/1000$ & Stochastic & $\mathcal{N}(0,\;3)$  \\
        \hline
    \end{tabular}
    \caption{\label{tab:priors} Set of cosmological, axion, bias, and counterterm parameters with their corresponding priors. $\mathcal{U}(a,b)$ indicates that the prior is a uniform distribution with lower and upper bounds $a$ and $b$ and $\mathcal{N}(\mu,\sigma)$ indicates that the prior is a Gaussian distribution with mean $\mu$ and variance $\sigma^2$. The parameter $h$ is the Hubble constant today in units of $100$ km/s/Mpc and the stochastic terms are quoted in units of $[\mathrm{Mpc}/h]^3$.}
\end{table}

As our dataset, we use the CMASS combined NGC and SGC as well as the LOWZ combined NGC and SGC (made public by the BOSS collaboration~\cite{Gil-Marin2016TheGalaxies}). The covariance matrices are computed from the MultiDark-Patchy mock catalogs \cite{Kitaura2016TheRelease,Rodriguez-Torres2016TheRelease}. These simulations were run using an augmented Lagrangian perturbation theory algorithm combined with a bias calculation scheme to populate the simulated halos with galaxies~\cite{Kitaura2014ModellingBiasing}. We fit the monopole ($\ell=0$) and quadrupole ($\ell=2$) moments of the anisotropic galaxy power spectrum as defined in Eq.~(\ref{eq:P_ell}).
To calculate the posterior distribution for the parameters of Table~\ref{tab:priors}, we use the MCMC sampler \texttt{emcee}~\cite{Foreman-Mackey2013emcee} with which we create our own likelihood code adapted for axion parameters. The priors on the cosmological parameters are chosen to match closely those of similar analyses and are generic for $\Lambda$CDM parameters estimation using BOSS data~\cite{Ivanov2020CosmologicalParameters}. The priors on the nuisance parameters are theoretical priors motivated by the fact that EFT parameters are expected to be $\mathcal{O}(1)$~\cite{Chudaykin2020NonlinearPerturbation,Nishimichi2020BlindedChallenge}.

We employ the same fitting method as Ref.~\cite{Hlozek2015AData} to obtain valid $\Lambda$CDM + ultralight axions priors instead of the regular $\Lambda$CDM priors from the Planck analysis. We check fo MCMC convergence using the spectral method of Ref~\cite{Dunkley2005FastAnd}. 
Finally, we compile all of the posterior distributions for the axion fraction as a function of the axion mass for the galaxies-only and galaxies+CMB scenarios and obtain a \textit{constraint plot} shown in Fig.~\ref{fig:combined_constraint} for the $m_a-\Omega_a/\Omega_d$. 


\section{Tests on Simulations}\label{sec:simtest}
We test our model on two sets of simulated data. In the first, we use the MultiDark-Patchy Mocks~\cite{Kitaura2016TheRelease,Rodriguez-Torres2016TheRelease} set of simulations provided with the BOSS datasets. These are the standard simulations used to validate galaxy clustering models in redshift space. We run our model on the monopoles and quadrupoles extracted from the simulated CMASS NGC data in order to compare the posterior distribution of our cosmological parameters with the true values used in the simulations. For the second set, we use the large scale Lagrangian based simulation code \texttt{Peak-Patch}~\cite{Stein2019TheValidation} in order to generate our own set of simulations. We introduce axions in one set of simulations and assume $\Lambda$CDM conditions for the other. We extract the multipoles from the simulations before running our model on this set of simulations as well. This is to determine if axionic components in the dark matter can be revealed by our model if they exist.

\subsection{MultiDark-Patchy Mocks}\label{sec:patchyNGC}
In order to validate the model presented in Sec~\ref{sec:model}, we make use of the MultiDark-Patchy mock simulations. These simulations are based on a $\Lambda$CDM cosmology and do not include ultralight axions. We use the priors in Table~\ref{tab:priors} and we test using two priors on $\omega_b$, with one from the CMB and one from BBN. Using these two choices of priors, we make two analyses with the axion fraction fixed to zero (assuming pure CDM) and two analyses allowing the axion fraction to vary (assuming an axion mass of $m_a=10^{-27}$ eV), for a total of four runs. We also set the sum of the neutrino masses to zero since the simulations were run without massive neutrinos.
\begin{figure}
    \centering
    \begin{subfigure}[t]{0.6\textwidth}
        \includegraphics[width=1\linewidth]{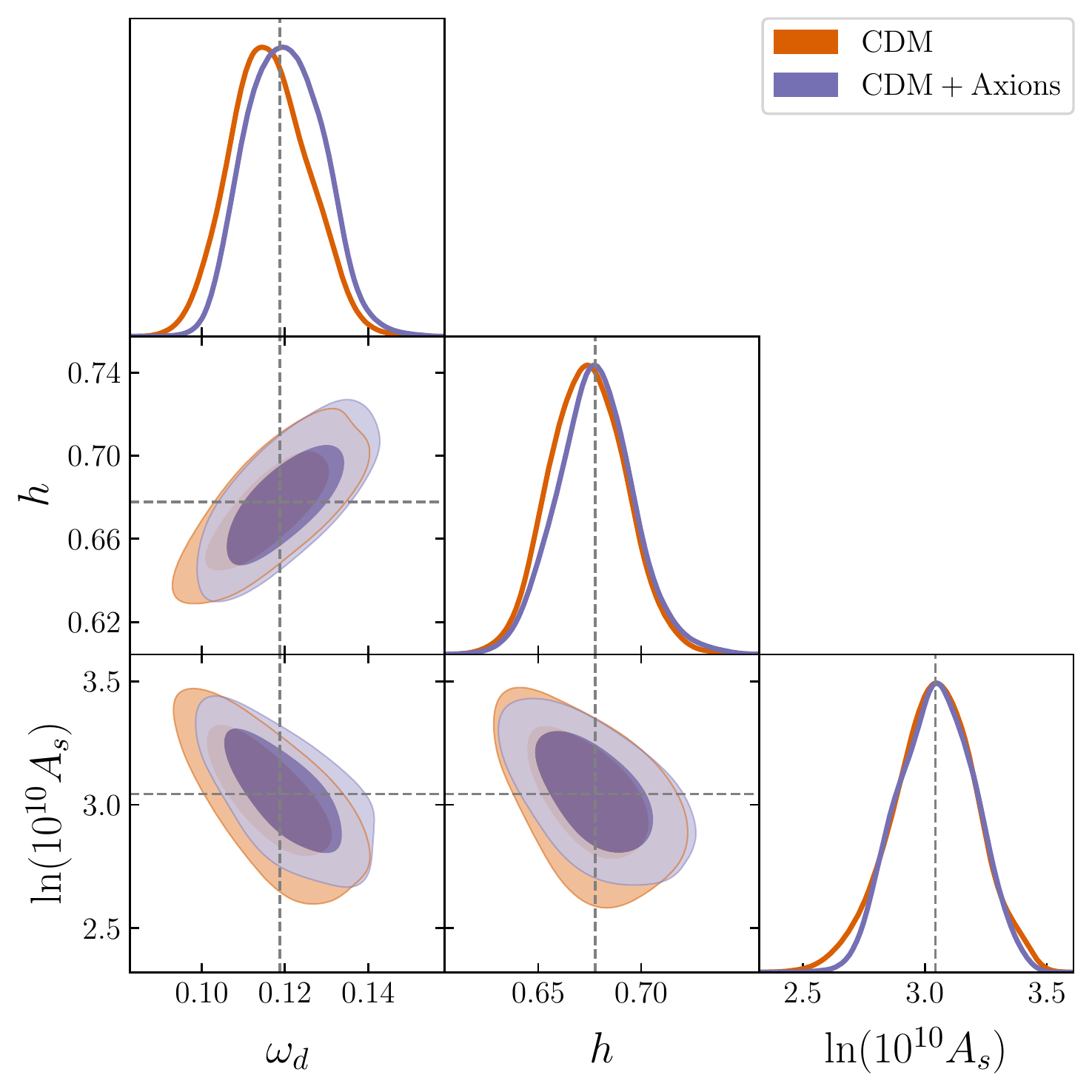}
        \caption{}
        \label{fig:triangle_Planck_omegab_Mocks}
    \end{subfigure}
    ~
    \begin{subfigure}[b]{0.6\textwidth}
        \includegraphics[width=1\linewidth]{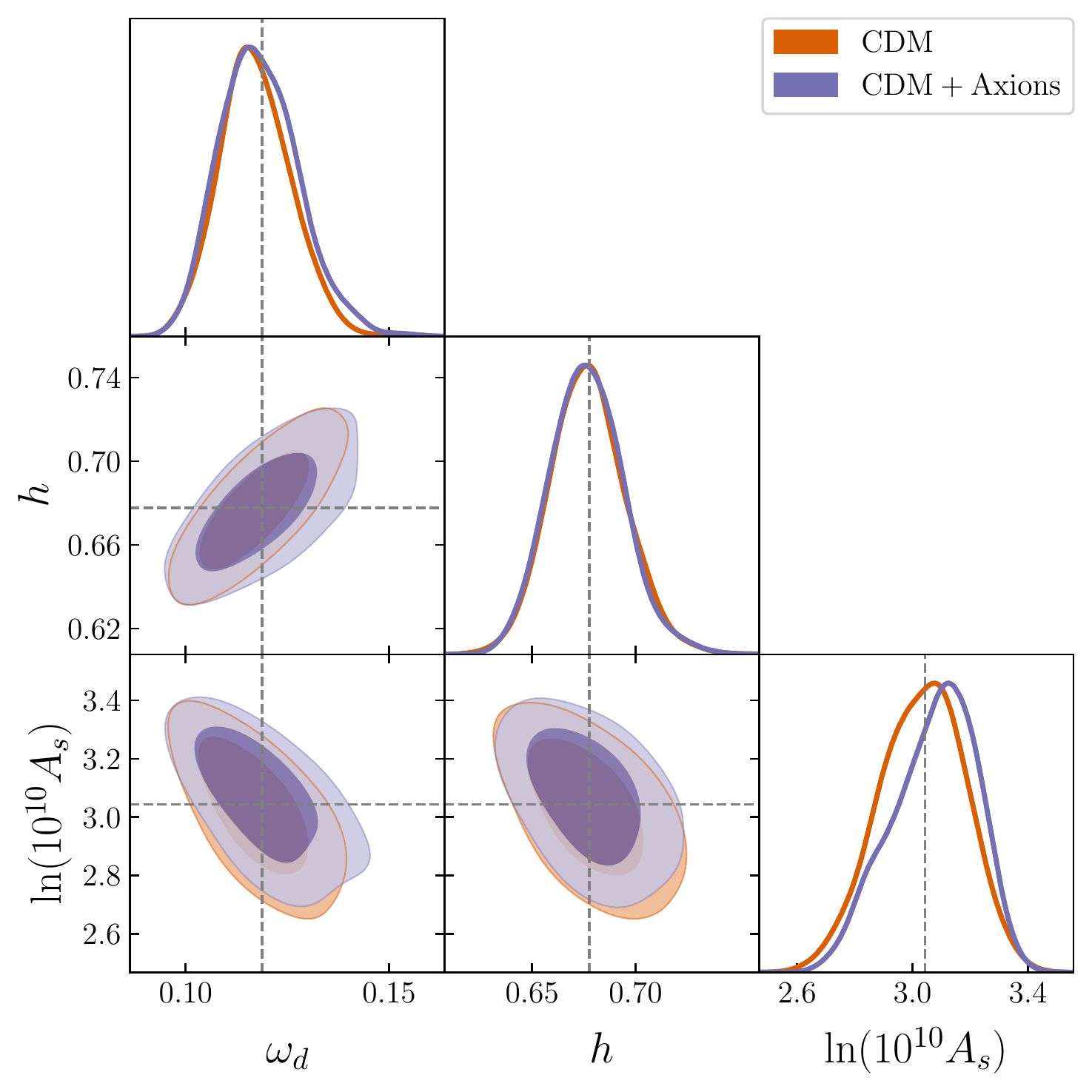}
        \caption{}
        \label{fig:triangle_BBN_omegab_Mocks}
    \end{subfigure}
    \caption{Model validation on high-$z$ NGC mock catalogs with a Planck (a) or BBN (b) prior on $\omega_b$. The dashed lines denote the input parameters of the simulations. CDM denotes the use of the $\Lambda$CDM model while the CDM + Axions denotes the extended CDM model where we also vary the axion fraction. The NGC mocks were run with a CDM cosmology.}
\end{figure}

\begin{table}
    \centering
    \begin{tabular}{ |c|cc|cc| }
    \hline
     $\omega_b$ Prior & \multicolumn{2}{c|}{Planck Prior} & \multicolumn{2}{c|}{BBN Prior}  \\
    \hline
    \hline
    Parameter & Best-Fit & Mean \& 68\% C.L. & Best-Fit & Mean \& 68\% C.L. \\
    \hline
    $100 \omega_b$ & $2.218$ & $2.215 \pm 0.015$ & $2.220$ & $2.215\pm 0.038$ \\
    $\omega_d$ & $0.110$ & $0.1171^{+0.0082}_{-0.0094}$ & $0.118$  & $0.1163^{+0.0089}_{-0.010} $\\
    $\ln(10^{10} A_s)$ & $3.171$ & $3.03\pm 0.15$ & $3.055$ &$3.04^{+0.18}_{-0.16}$ \\
    $h$ & $0.665$ & $0.677^{+0.017}_{-0.019}$ & $0.676$ &$0.674\pm 0.019$ \\
    $b_1$ & $1.828$ & $1.97^{+0.15}_{-0.18}$ & $1.928$ &$1.96^{+0.16}_{-0.21}$ \\
    $b_2$ & $0.345$ & $0.59^{+0.23}_{-0.43}$ & $0.442$ & $0.53^{+0.27}_{-0.44}$ \\
    $b_3$ & $-1.667$ & $-0.7\pm 1.8$ & $-0.659$ & $-0.8^{+1.8}_{-1.6}$ \\
    $c_\mathrm{ct}$ & $-0.725$ & $-0.4^{+1.9}_{-2.1}$ & $0.615$ &$-0.4\pm 2.0$ \\
    $\tilde{c}_{r,1}$ & $-13.445$ & $-16.3^{+6.9}_{-4.4}$ & $-14.903$ & $-15.4^{+6.1}_{-5.2}$\\
    $\tilde{c}_{\epsilon,1}/1000$ & $3.942$ & $2.7^{+1.8}_{-2.7}$ & $2.910$ &$3.0^{+2.0}_{-2.9}$ \\
    $\tilde{c}_{\epsilon,2}/1000$ & $-0.640$ & $-0.8\pm 2.4$  & $-0.659$ &$-0.5\pm 2.5$ \\
    \hline
    $\Omega_m$ & $0.300$ & $0.304^{+0.012}_{-0.013}$ & $0.306$ & $0.304^{+0.012}_{-0.015}$ \\
    $\sigma_8$ & $0.830$ & $0.806\pm 0.046$ & $0.821$ &$0.804^{+0.052}_{-0.042}$ \\
    \hline
    \end{tabular}
    \caption{Marginalized maximal likelihood, mean and $1\sigma$ bounds on the direct and derived parameters in our analysis from the mock high-$z$ NGC catalogs. The results are for the pure CDM runs with a prior on $\omega_b$ taken from Planck (left) or BBN (right). The units are the same as for Table~\ref{tab:priors}.}
    \label{tab:NGC_no_axions_Planck}
\end{table}

The marginalized posterior for $h$,  $\ln(10^{10} A_s)$, and $\omega_d$ are given in Fig.~\ref{fig:triangle_Planck_omegab_Mocks} for two values of the prior on $\omega_b$ and with the inclusion or exclusion of axions. The full results for all parameters including the derived $\Omega_m$ and $\sigma_8$ are given in Table~\ref{tab:NGC_no_axions_Planck} for the axion-free runs. The results for the latter tests using the BBN prior are compatible with analyses run for the \texttt{PyBird} code~\cite{DAmico2020LimitsOn}. The main difference we have observed is that our estimated value of $A_s$ is unbiased contrary to the $2.3\sigma$ tension found in the original \texttt{PyBird} analysis. We attribute this to the only difference which is in our treatment of the stochastic parameters (see Appendix~\ref{app:perturbations}). We specifically include stochastic contributions to the monopole and find that the inclusion of a stochastic term for the quadrupole does not affect our results. We note finally that we do not observe the aforementioned bias in our analysis of the BOSS DR12 data using $\Lambda$CDM.


\subsection{Peak-Patch Mocks}\label{sec:peakpatchmocks}

\begin{figure}
    \centering
    \begin{subfigure}[l]{0.5\textwidth}
        \centering 
        \includegraphics[height=0.21\textheight]{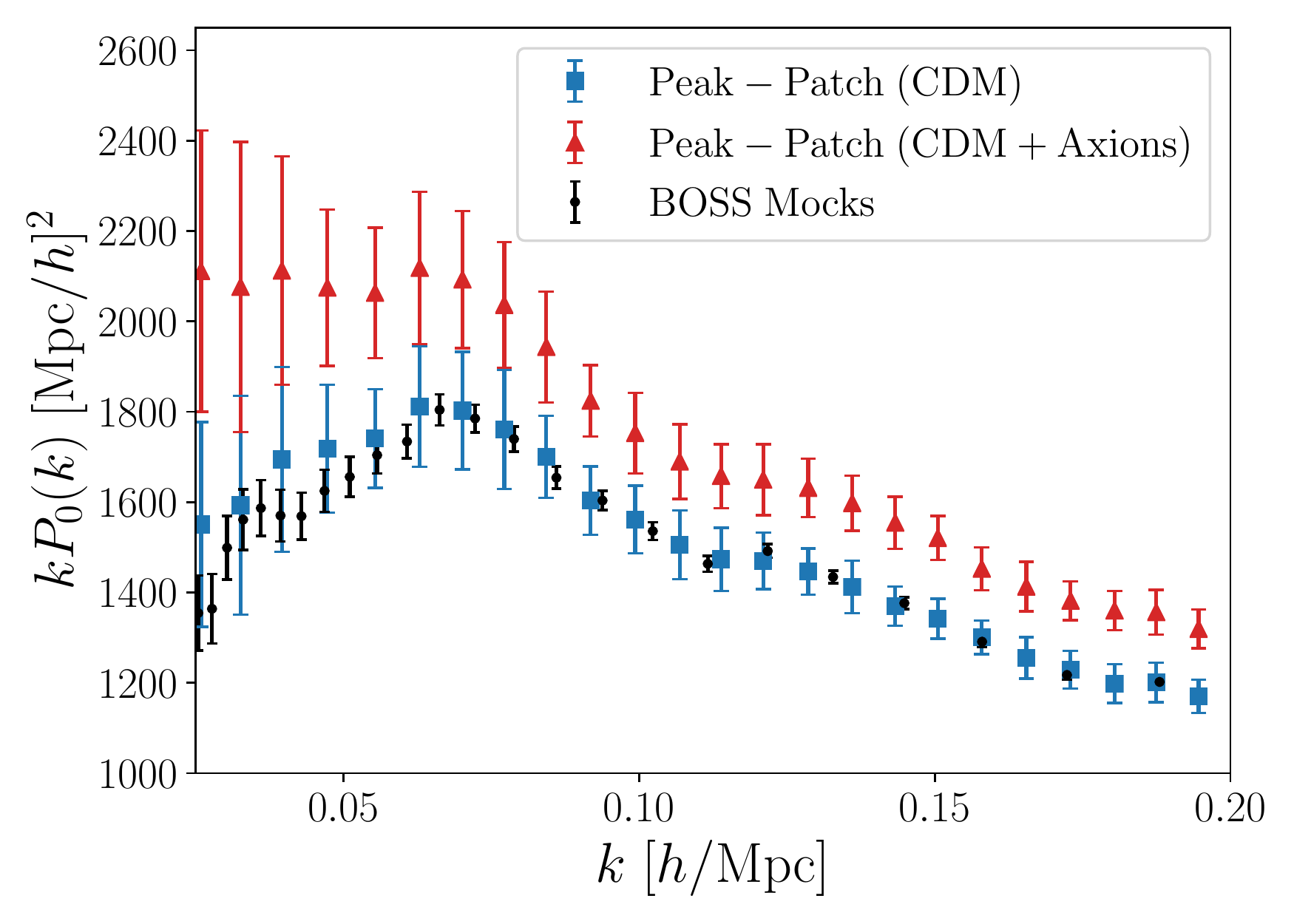}
        \caption{\label{fig:PP_mocks_mon}}
    \end{subfigure}%
    ~
    \begin{subfigure}[r]{0.5\textwidth}
        \centering 
        \includegraphics[height=0.21\textheight]{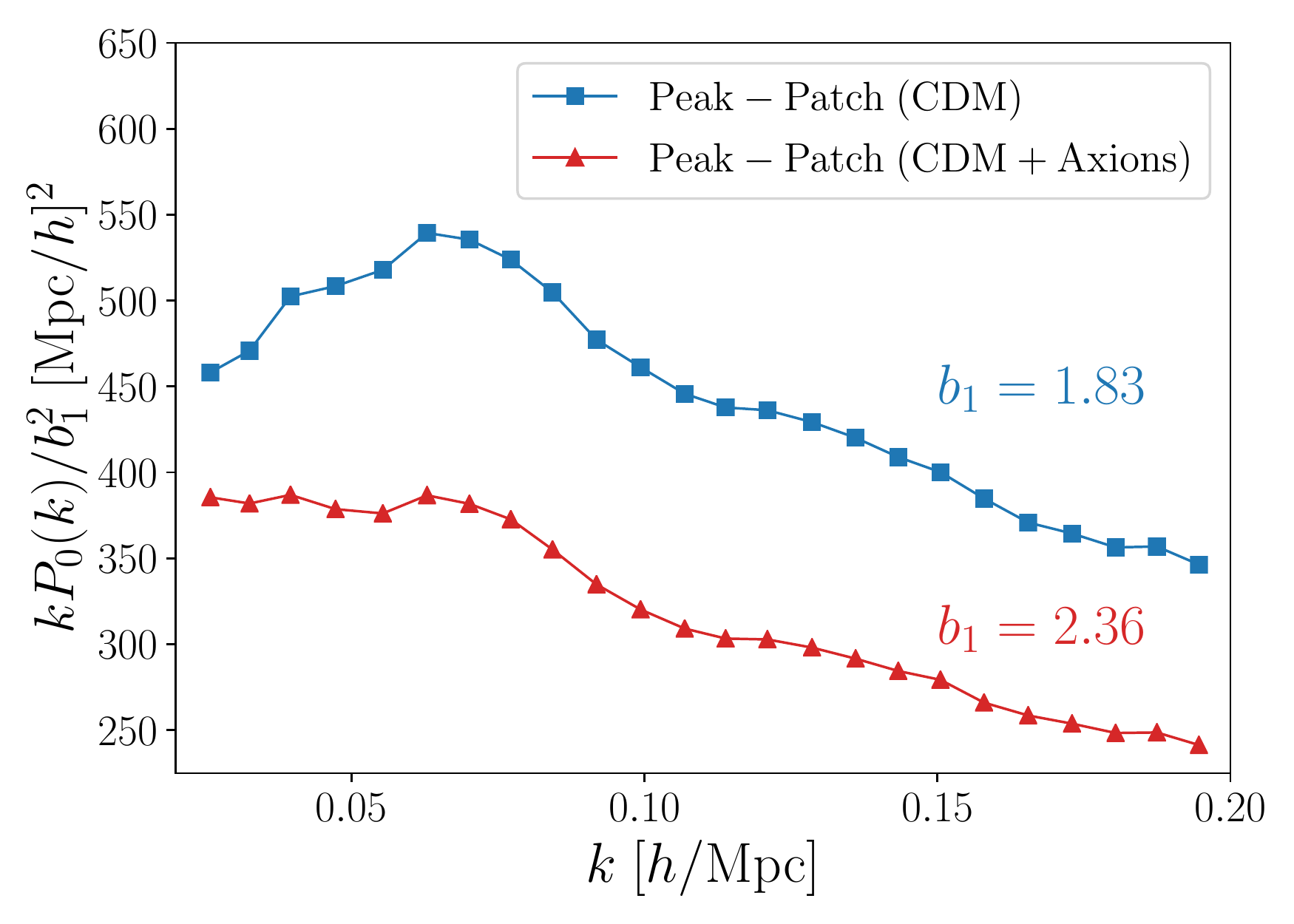}
        \caption{\label{fig:PP_mocks_bias}}
    \end{subfigure}
    ~
    \begin{subfigure}[r]{0.5\textwidth}
        \centering 
        \includegraphics[height=0.21\textheight]{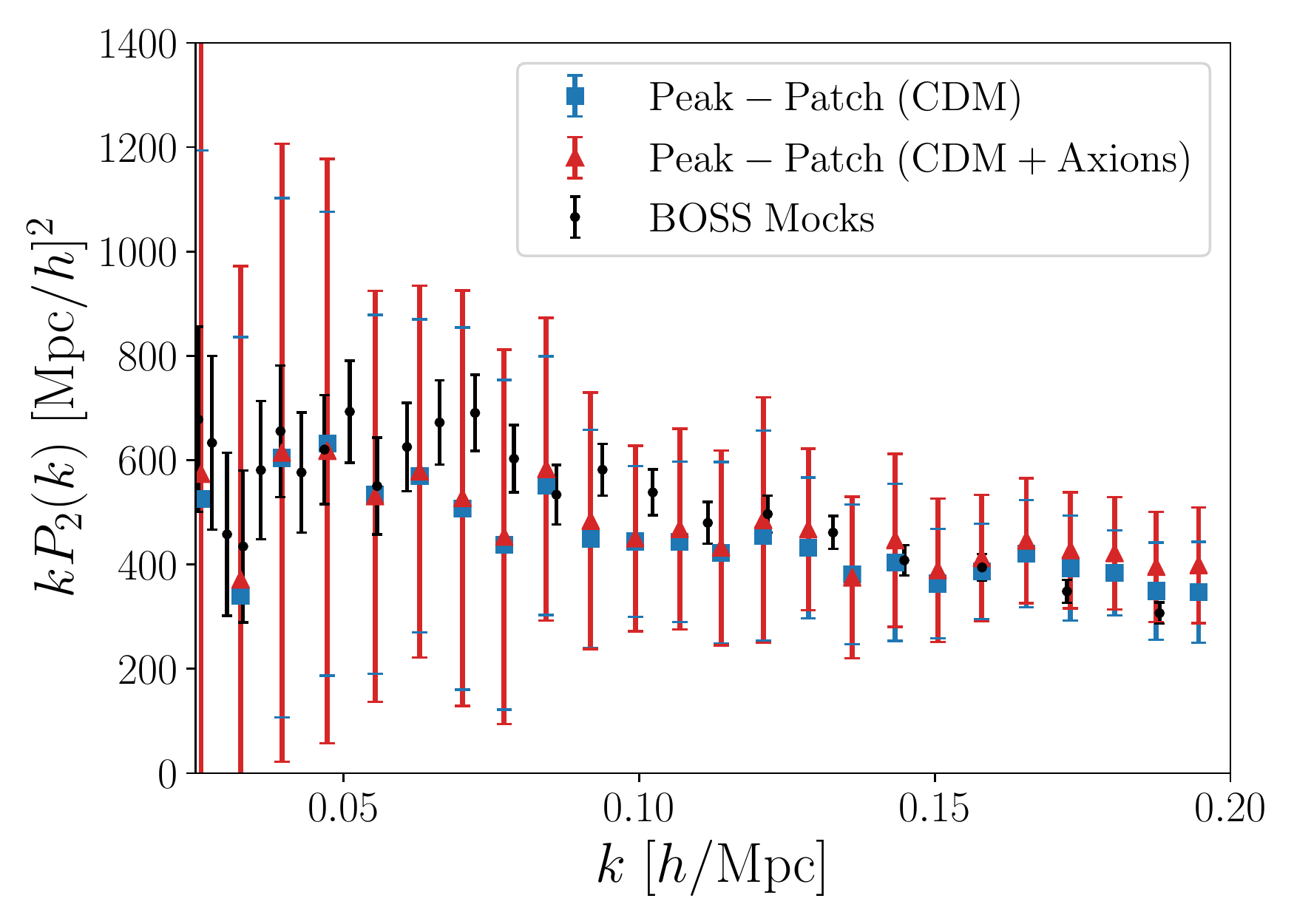}
        \caption{\label{fig:PP_mocks_quad}}
    \end{subfigure}%
    ~
    \begin{subfigure}[r]{0.5\textwidth}
        \centering 
        \includegraphics[height=0.21\textheight]{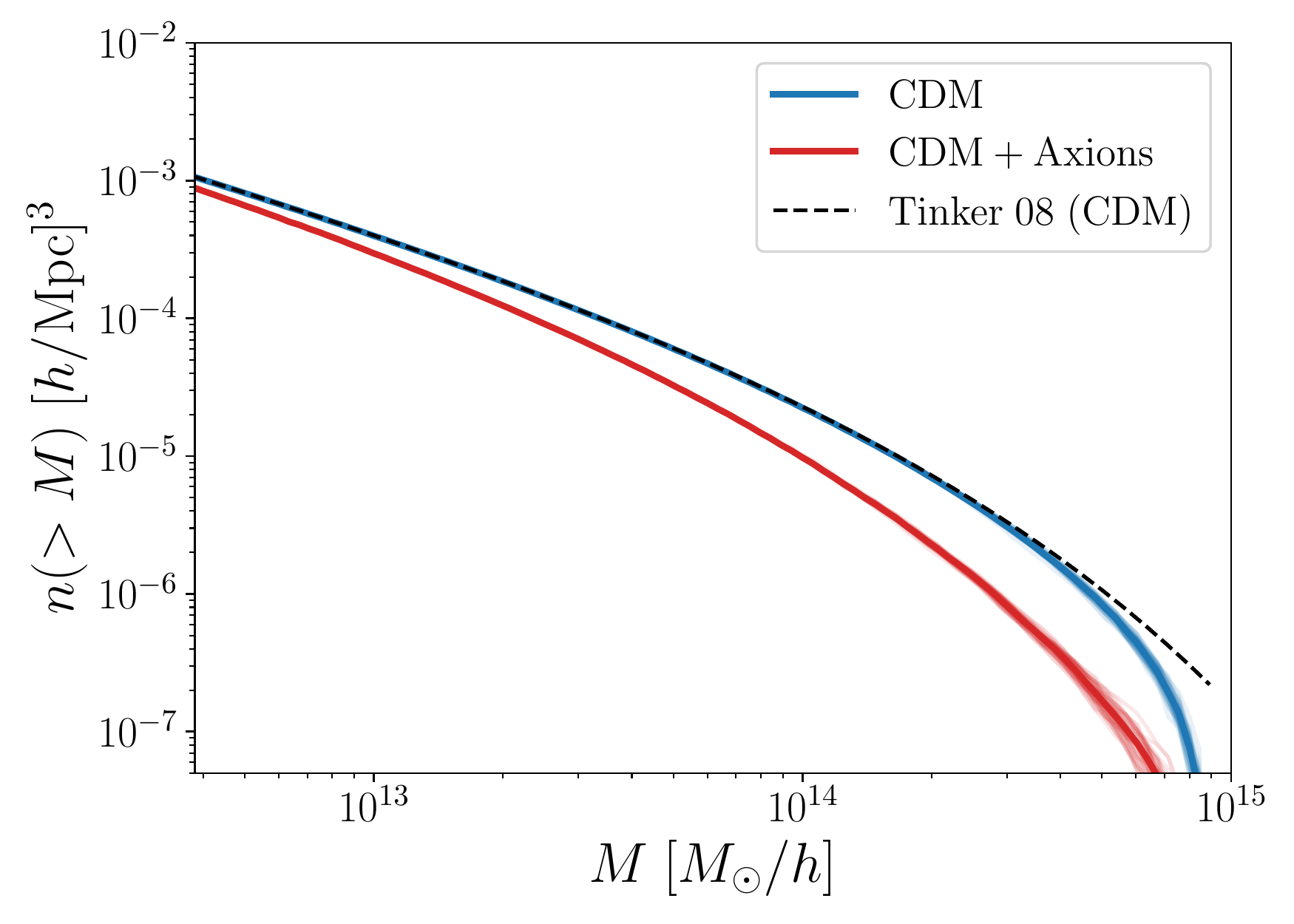}
        \caption{\label{fig:PP_mocks_hmf}}
    \end{subfigure}
    \caption{Analysis of the Peak-Patch mock galaxy catalogs where the CDM scenario denotes an axion-free cosmology and where CDM+Axions denotes a 5\% concentration of $10^{-28}$ eV axions. Such high axion fraction is disfavoured by observations and is meant to serve as an illustrative case. (a) Simulated galaxy monopoles with the BOSS simulated catalogs results used for HOD calibration, (b) galaxy monopoles when removing the contribution of the linear bias showing a closer similarity with the linear results shown in Fig.~\ref{fig:linear_pk} with the error bars removed for clarity, (c) same as (a) for the quadrupoles, (d) halo mass functions of the halos in the simulation boxes with the Tinker result used for the CDM calibration.}
    \label{fig:PP_mocks}
\end{figure}

\begin{figure}
    \centering
    \includegraphics[width=0.8\linewidth]{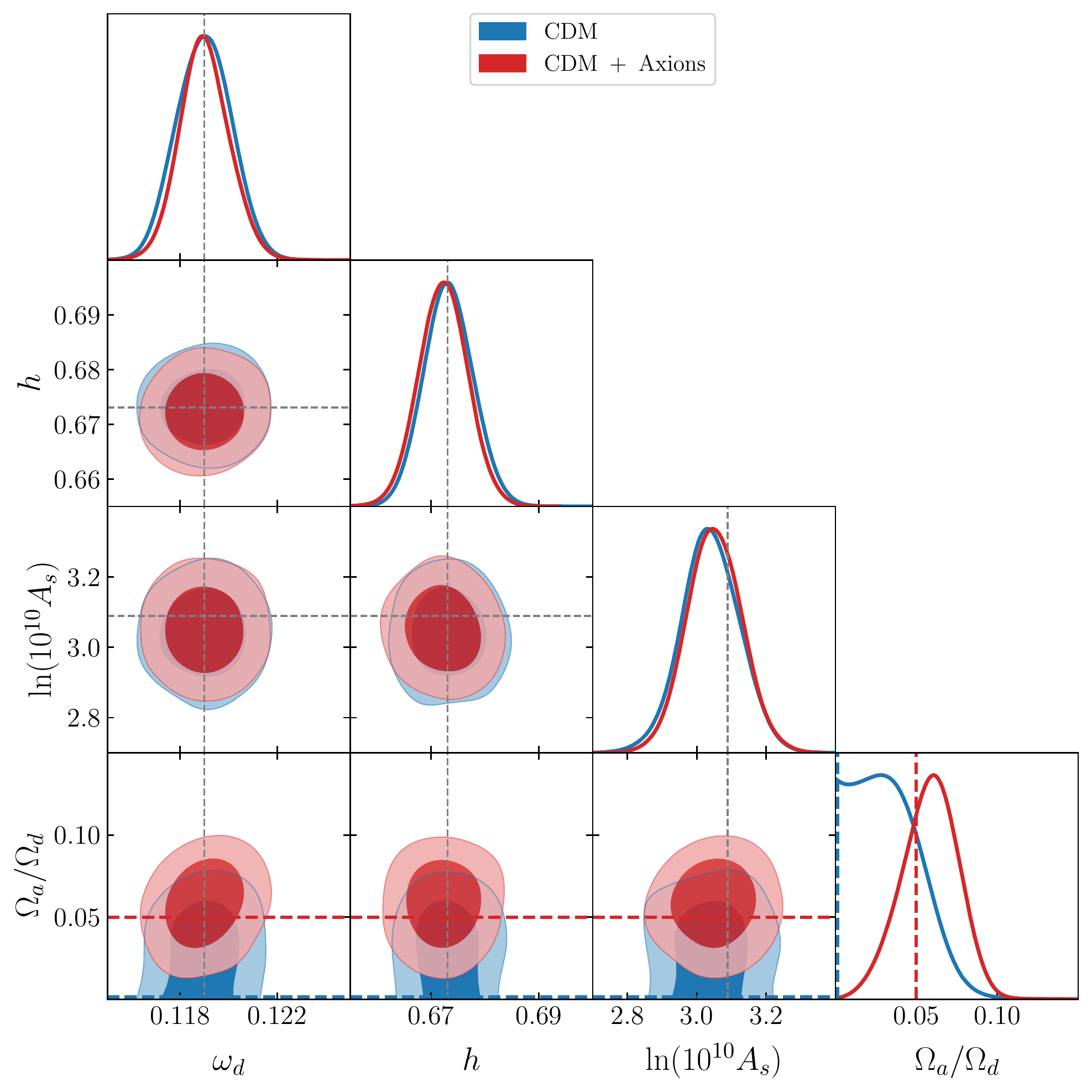}
    \caption{Posterior distributions of the cosmological parameters, including the axion density for the Peak-Patch mock galaxy catalogs. The covariance matrix is taken from the MultiDark-Patchy mock simulations and scaled by 16 with a Planck prior on $\{\omega_d, \omega_b, \ln(10^{10}A_s), h\}$. The CDM+Axions scenario denotes simulations containing 5\% of $10^{-28}$ eV axions.}
    \label{fig:PP_triangle_plot}
\end{figure}

We generate a set of 88 (1024 Mpc)${}^3$ boxes at a grid resolution of 1 Mpc. The boxes have a mean redshift of $z\approx 0.57$, and half of them are based on a $\Lambda$CDM cosmology with parameters $\{h=0.67, \ln 10^{10} A_s= 3.091, \omega_b=0.02222, \omega_d=0.119\}$ and the other half with the same cosmological parameters but with a non-zero axion concentration of $\Omega_a/\Omega_d=0.05$ for an axion mass of $m_a = 10^{-28}$ eV. We use the large-scale structure simulation code \texttt{Peak-Patch}~\cite{Stein2019TheValidation} which finds collapsed halos from a density field on a uniform grid and displaces them to their final positions using second order Lagrangian perturbation theory. The algorithm has been shown to reproduce very closely the results of $N$-body simulations. To generate realistic halo catalogs, we only allow the CDM component to cluster into halos given that the Jeans scale of axions at $10^{-28}$ eV forbids them to cluster even at later times. This lack of clustering combined with the suppression of power on most scales in the simulation lead to a strong deficit in very massive halos. This is again mirrored by results involving massive neutrinos where $N$-body simulations show the same type of halo mass function (HMF) difference as that obtained in our simulations~\cite{Castorina2013CosmologyWM,Castorina2015DEMNUni}. The HMF obtained in the presence of ultralight axions is shown in Fig.~\ref{fig:PP_mocks_hmf} also agrees with the analytical result of Ref.~\cite{Bauer2021IntensityMapping} for the same axion mass.

We then populate the halos with galaxies using a halo occupation distribution (HOD) technique. We use a similar prescription to that employed in Refs.~\cite{Tinker2011Cosmological, Beutler2014Clustering} where the expectation values of the number of central and satellite galaxies for a halo of mass $M$ are given by 
\begin{align}
    \left\langle N_{\text {cen }}\right\rangle_{M}&=\frac{1}{2}\left[1+\operatorname{erf}\left(\frac{\log M-\log M_{\min }}{\sigma_{\log M}}\right)\right],
    \\\left\langle N_{\text {sat }}\right\rangle_{M}&=\left\langle N_{\text {cen }}\right\rangle_{M}\left(\frac{M}{M_{\text {sat }}}\right)^{\alpha} \exp \left(\frac{-M_{\text {cut }}}{M}\right).
\end{align}
We set $M_\mathrm{min}=10^{13.09}\;M_\odot/h$, $\sigma_{\log M}=0.19$, $M_\mathrm{cut}=10^{14.2}\;M_\odot/h$, $M_\mathrm{sat}=10^{12.9}\;M_\odot/h$, $\alpha=0.25$. These values are obtained by computing the galaxy multipoles for the HOD parameters over a 6-dimensional grid near the values found in the literature. The final parameters are chosen to minimize the $\chi^2$ between the BOSS mock galaxy catalog multipoles and the average of the CDM-only simulations. For the BOSS mocks, we use the CMASS sample given the redshift of our simulated boxes.

As our goal is to test our model in a simulated setting, we do not reproduce all of the characteristics of the BOSS survey. For instance, we do not account for the limited survey window. To remain self-consistent, we do not apply a window function when running our MCMC analysis with this dataset. The galaxy multipoles for the simulated boxes are shown in the two top panels of Fig~\ref{fig:PP_mocks}. We can see that the approximate fitting procedure results in a very close monopole to the BOSS simulations. The quadrupole is not as well reproduced but still gives a reasonable approximation on large scales. The small-scale discrepancy may be due to attribution of velocities to satellite galaxies. The amplitude of the monopoles initially seems to contradict the linear predictions, but Fig.~\ref{fig:PP_mocks_bias} shows that the expected suppression of power due to axions is recovered when taking into account the biasing effects of galaxies. The value of the bias used is the best-fit value from our statistical analysis of the galaxy multipoles with our EFTofLSS model.

Given that our number of simulations is much smaller than the $\sim 1000$ of the BOSS simulated dataset, we use the MultiDark-Patchy mocks covariance matrix scaled down by a factor of 16. We also limit the maximal scale to $k=0.2\;h/$Mpc so that we have a reliable modelling of the quadrupole. The main impediment in designing these simulations is that mock galaxy catalogs are calibrated on $N$-body simulations with galaxy formation. In our case, such a calibration was not possible since a mixed axion dark matter simulation suite with baryons is not yet available. We assume that the HOD parameters are unchanged from the pure CDM case when including a 5\% component of ultralight axions. The study of the impact of subdominant ultralight axionic components on galaxy formation will be left for future studies once the computational tools have been developed. For this reason, we prefer to keep the error margin on the monopole and quadrupole moments for these simulations high, to use these results conservatively, and to impose Planck-like priors to all cosmological parameters to isolate the effects of axions.

A supplementary application of our simulations is to investigate what kind of biases the existence of axions would bring in the measurements of cosmological parameters. Since we impose strong priors on the varied parameters and since the covariance is large, we do not observe any bias. However, derived and nuisance parameters may be biased even with CMB priors. 
\begin{figure}
    \centering
    \includegraphics[width=0.45\linewidth]{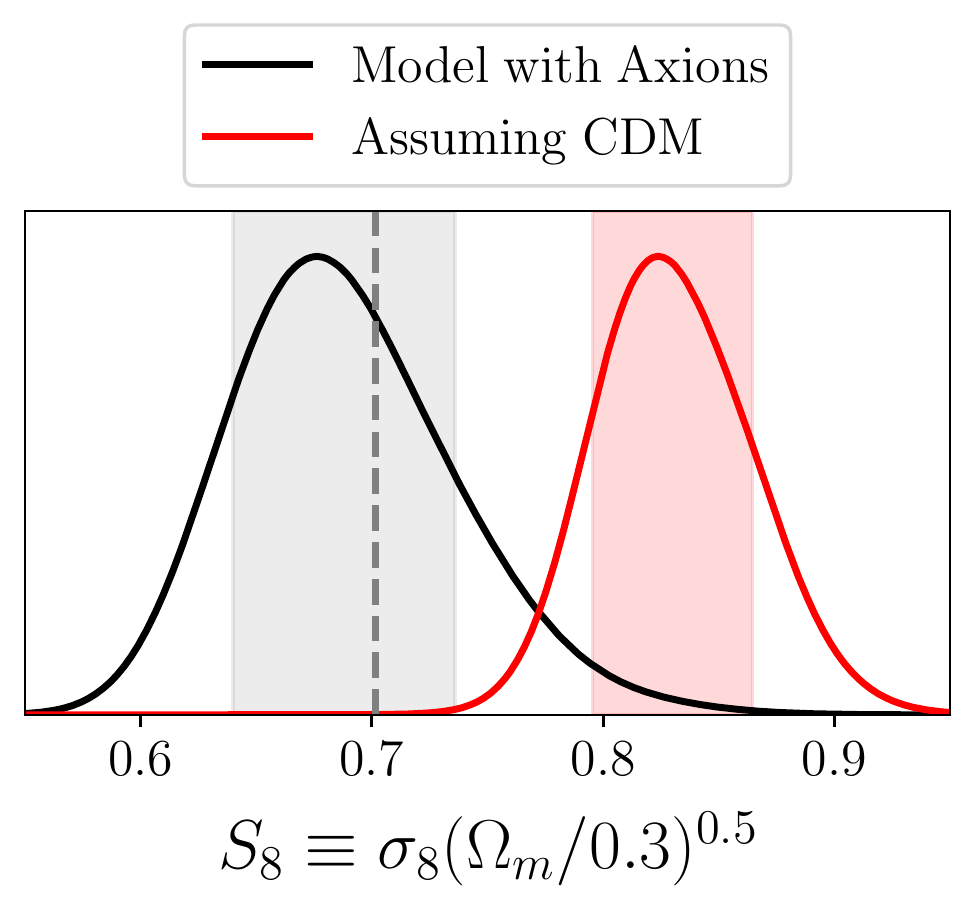}
    \caption{Marginalised posterior distributions of the $S_8$ parameter measured for a simulated universe with 5\% of $10^{-28}$ eV axions. We observe that assuming that dark matter is purely CDM in a universe containing axions leads to a biased measurement by many standard deviations. The $1\sigma$ intervals for each of the posteriors are highlighted by shaded areas with the true value of $S_8$ denoted by the dashed line. $S_8$ is very sensitive to power spectrum suppression making it a an ideal probe of ultralight axions.}
    \label{fig:sigma8_with_axions}
\end{figure}
To illustrate this, in Fig.~\ref{fig:sigma8_with_axions}, we show the posterior distribution of the $S_8\equiv \sigma_8 (\Omega_m/0.3)^{0.5}$ parameter which is a measure of large-scale structure clustering. We observe that the distribution for the total matter density ($\Omega_m$) is unchanged due to strong priors, but obtain a biased estimate on the derived $S_8$. We find that using a model assuming no axions in a universe where they exist as a subdominant component leads to biased measurements. The model accounts for the lack of clustering (lower $S_8$) by boosting the value of the linear galaxy bias. Also, a model with axions biases $S_8$ in galaxies high compared to the true value if you analyse assuming CDM. However, the effect is driven by the galaxy bias, which causes axions to increase the galaxy power spectrum relative to CDM. This enhancement is consistent with the same effect seen in the HI halo model~\cite{Bauer2021IntensityMapping}. Given our large covariance matrix, the $\chi^2$ difference between the two models is only about a factor of 2, but this is sufficient to obtain a lower Bayes information criterion (BIC)~\cite{Schwarz1978EstimatingThe} despite the addition of a free parameter. However the difference in BIC is too small to definitively favour one model over the other in this context. Better calibrated simulations with a more realistic covariance may shed more light on model selection in the presence on ultralight axions.

\section{Results}\label{sec:results}

\subsection{Galaxy Surveys Only}
\begin{figure}
    \centering
    \begin{subfigure}[l]{0.5\textwidth}
        \centering 
        \includegraphics[width=\linewidth]{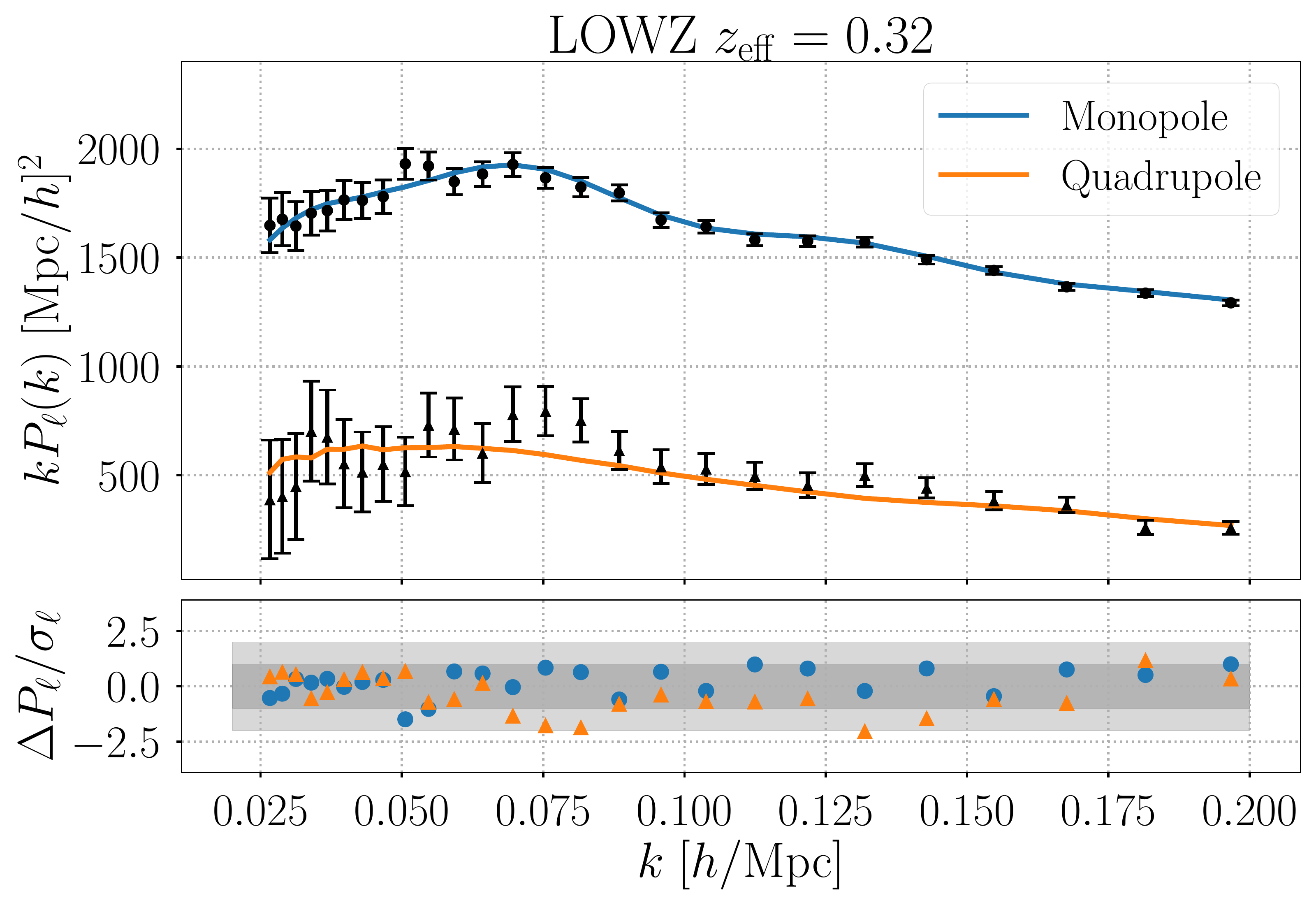}
        \caption{}
    \end{subfigure}%
    ~
    \begin{subfigure}[r]{0.5\textwidth}
        \centering 
        \includegraphics[width=\linewidth]{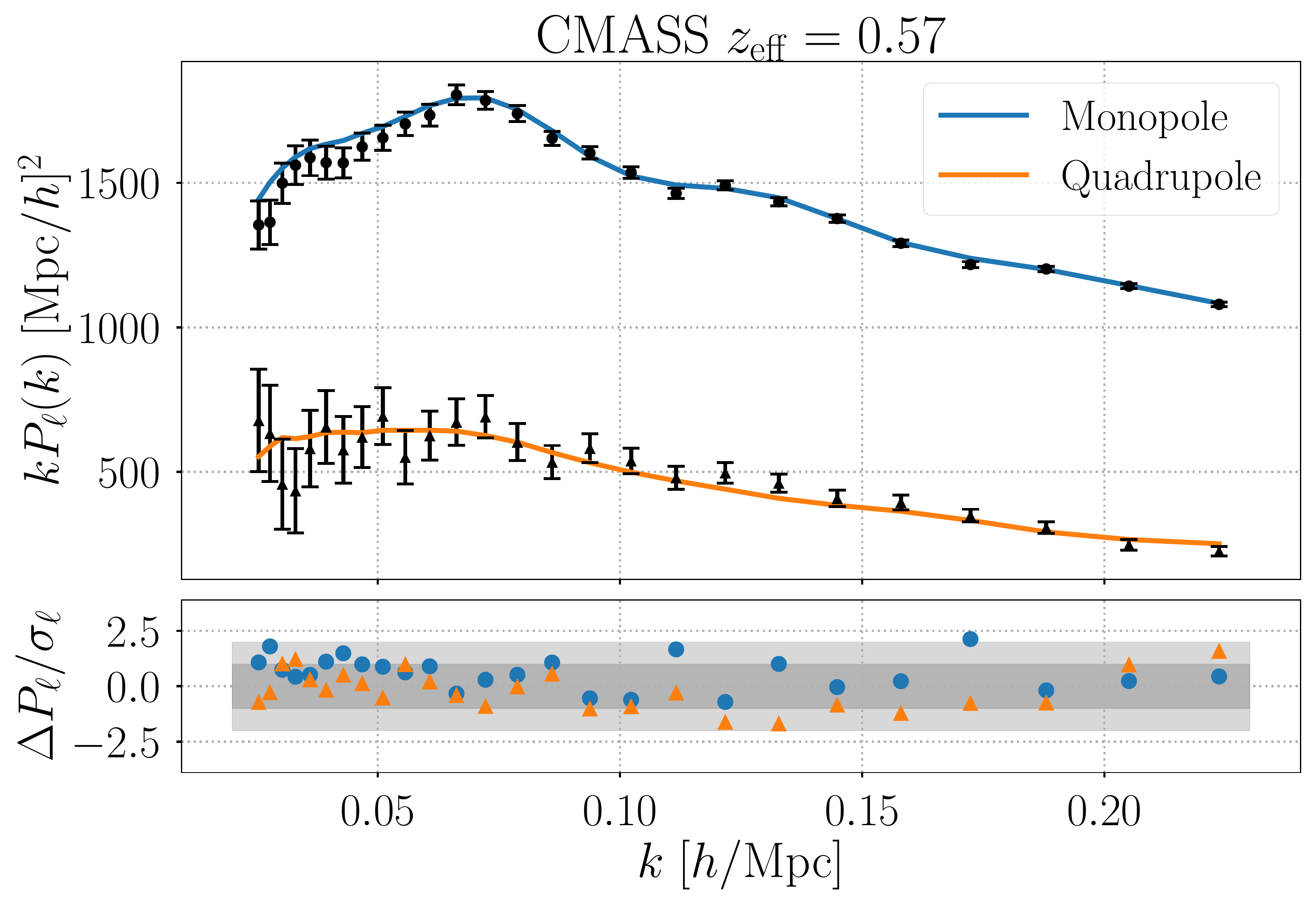}
        \caption{}
    \end{subfigure}
    \caption{Maximum likelihood curve of the monopole and quadrupole for the (a) LOWZ and (b) CMASS datasets. The model displayed has an axion mass of $m_a = 10^{-27}$ eV. The bottom panels show the residuals.}
    \label{fig:best_fit}
\end{figure}
\begin{figure}
    \centering
    \includegraphics[width=0.5\linewidth]{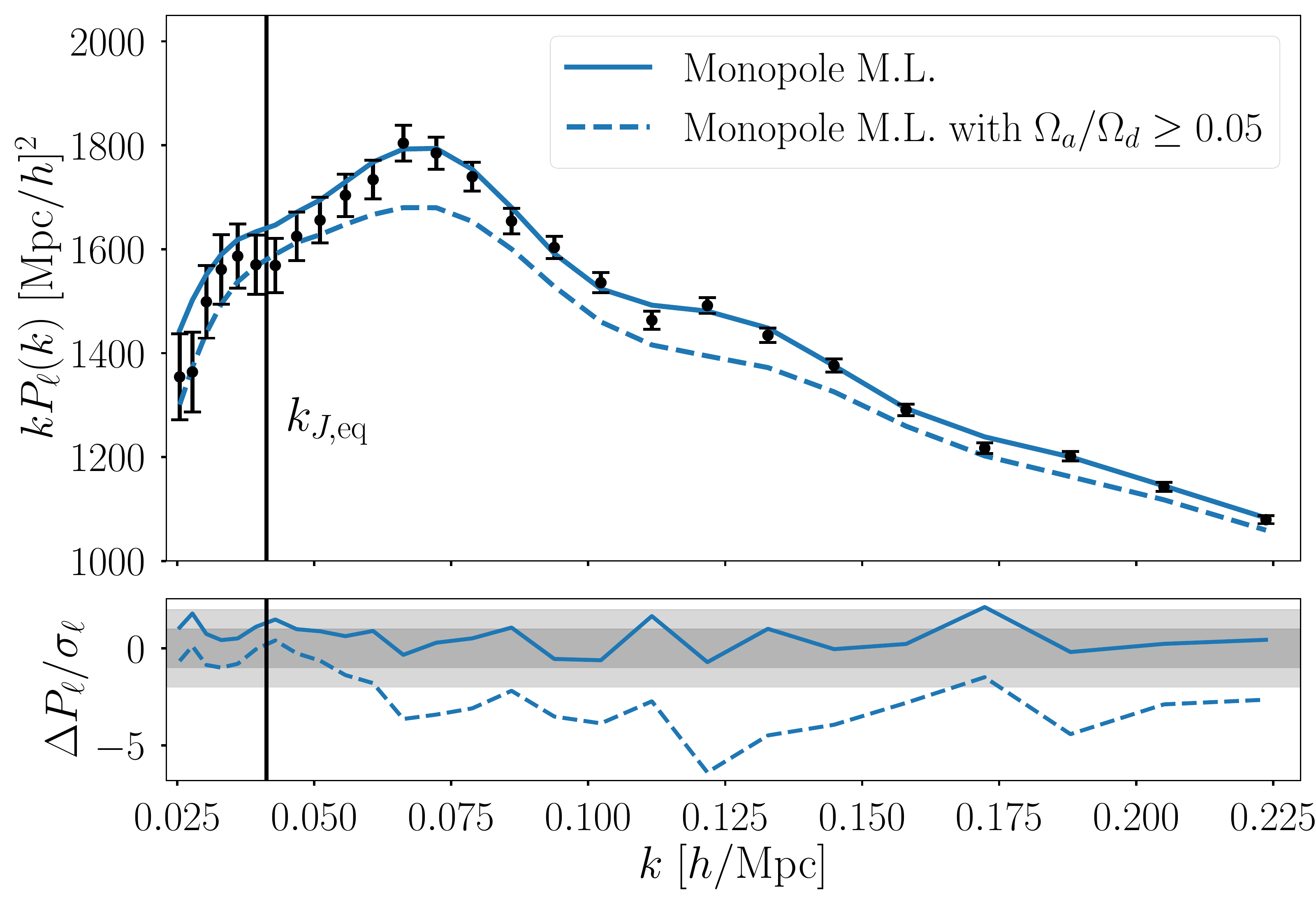}
    \caption{Maximum likelihood (M.L.) monopole for $m_a=10^{-27}$ eV. The full line denotes the overall maximum likelihood curve identical to the CMASS result of Fig.~\ref{fig:best_fit}. The dashed line denotes the maximum likelihood imposing that axions compose at least 5\% of the dark matter. We note that this value is about double the 95\% C.L. upper bound for ultralight axions from CMB data and is used for illustration purposes. We also highlight the axion Jeans scale as it is the point at which the model begins to deviate from the data. Even if the linear spectra do match on small scales, the galaxy power spectra do not converge for $k<k_{J,\mathrm{eq}}$ since the values of the nuisance parameters are different for both fits.}
    \label{fig:monopole_fax_005.pdf}
\end{figure}

Using the model described in Section~\ref{sec:model}, we fit the monopoles and quadrupoles of the CMASS and LOWZ datasets simultaneously. For each run, we allow the axion density to vary but fix the axion mass. The best-fit result for the case with $m_a=10^{-27}$ eV is plotted along with the full dataset and the residuals in Fig.~\ref{fig:best_fit}. From the fits, we obtain an upper bound on the axion density at the 68\% and 95\% confidence levels. We give an example of a value of the axion fraction in tension with the data in Fig.~\ref{fig:monopole_fax_005.pdf}. In this case, the axion fraction is set to 5\% and the model deviates from observations at scales below the Jeans scale of the axions. This illustrates how the linear power spectrum suppression (see Fig.~\ref{fig:linear_pk}) leads to detectable effects in the galaxy clustering multipoles allowing us to extract information about the axion density from these data.

The combined constraints as a function of mass are displayed in Fig.~\ref{fig:combined_constraint} where the allowed regions are highlighted in shades of red. Over-plotted is the CMB analysis for ultralight axions for the Planck 2015 data release taken from Ref.~\cite{Hlozek2018UsingMatter}. We first note that the constraints obtained from the galaxy surveys are entirely independent on the CMB data with the exception that the power spectrum tilt has been fixed to the Planck~\cite{Aghanim2020Planck2018} best-fit value of $n_s=0.9611$ throughout this work with the exception of the analysis of simulations which have been run with $n_s=0.9655$. This matches the approach taken by other BOSS data analyses using similar models~\cite{DAmico2020TheCosmological}. Also, the sum of the neutrino masses has been fixed to $\sum m_\nu = 0.06$ eV as in part of the analysis conducted in Ref.~\cite{Hlozek2018UsingMatter}. The prior for the baryon density was taken from BBN and the analysis in this section can therefore serve as a verification of CMB studies.

\begin{figure}
    \centering
    \begin{subfigure}[b]{0.45\textwidth}
        \centering
        \includegraphics[width=\linewidth]{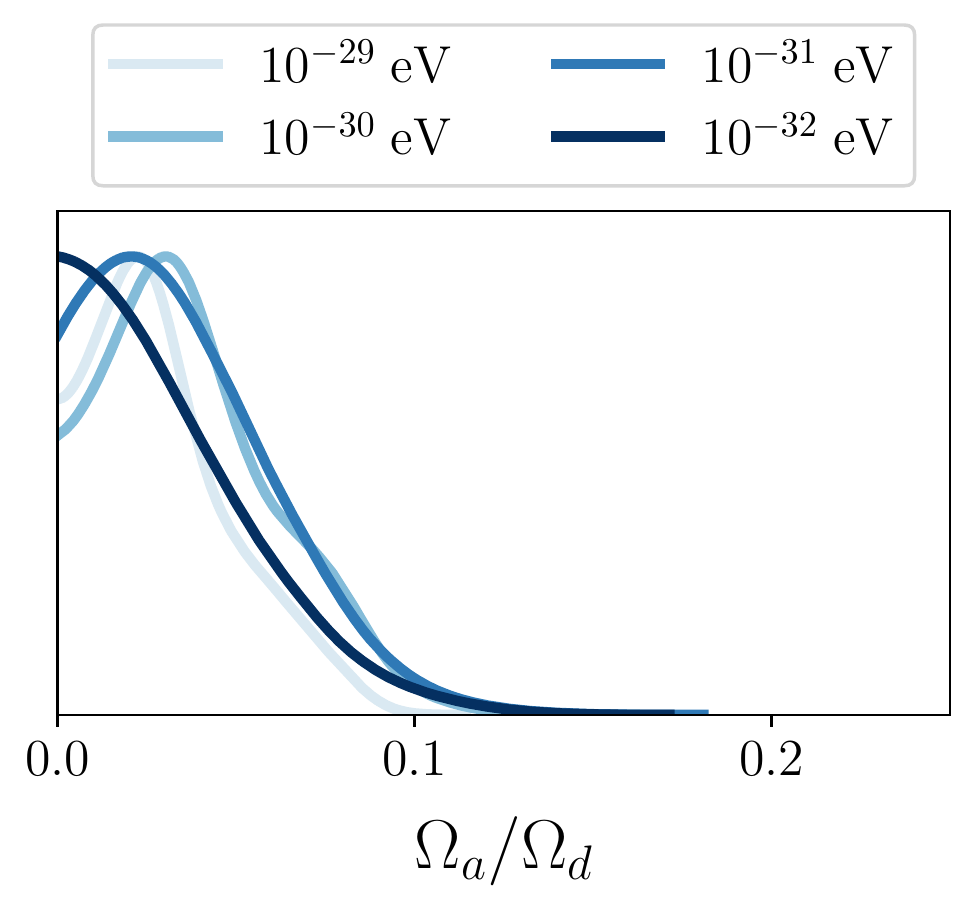}
        \caption{}
    \end{subfigure}%
    ~
    \begin{subfigure}[b]{0.45\textwidth}
        \centering
        \includegraphics[width=\linewidth]{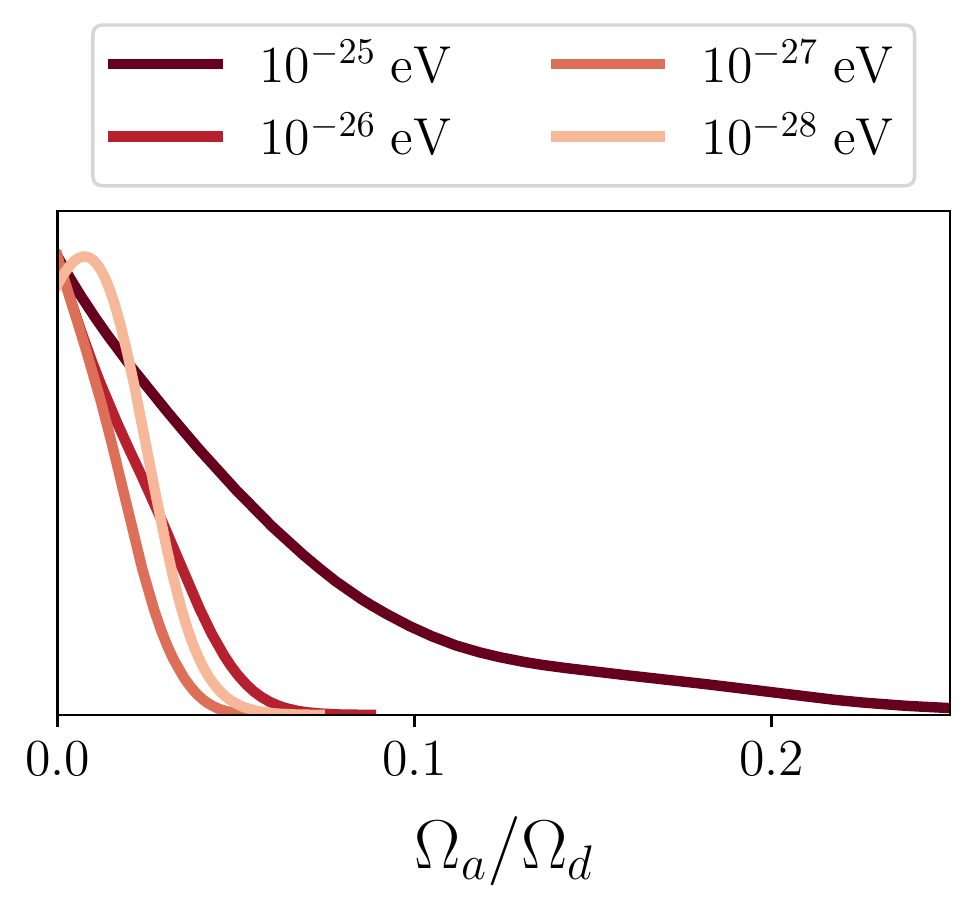}
        \caption{}
    \end{subfigure}
    \caption{Posterior distributions on the axion fraction for the range of axion masses studied. We note having much better constraint of the heavier axions $m_a\gtrsim 10^{-28}$ eV with the exception of $m_a= 10^{-25}$ eV where $k_{J,\mathrm{eq}} \approx k_\mathrm{max}$. The numerical values of the 95\% C.L. bounds from these marginalized posterior distributions are listed in Table~\ref{tab:2_sigma_bounds}. We note that only the axions with mass $m_a\gtrsim 10^{-27}$ eV on the right hand side are considered to be dark-matter-like since $m_a\lesssim H(a_\mathrm{osc})$.}
    \label{fig:posterior_axions}
\end{figure}

For masses below $10^{-28}$ eV, this study provides an important independent verification of the bounds on $\Omega_a$ found in previous work. Our $1\sigma$ exclusion bounds match those obtained from CMB very closely. In the heavier mass range, the constraints we obtain are improved by up to a factor of 3.7 relative to the CMB-only constraints. We show the full posterior distributions for the axion fraction in Fig.~\ref{fig:posterior_axions}. There we see more clearly the difference between the highest and lowest masses considered and we observe the increase in variance for the masses at the lower end. This is due to the choice of range in scales of galaxy surveys and the axion mass range. Given any choice of lower $k$ for the dataset, all axion masses for which $k_{m}(m_a)\ll k_\mathrm{min}$ will be indistinguishable from a simple shift in amplitude in the power spectrum. The scales below the Jeans scale will be suppressed by an amount proportional to the axion fraction. However, if all scales probed by the galaxy survey fall far below this scale, then all modes are suppressed by the same value. In this setting the axion fraction is degenerate with the primordial power spectrum amplitude ($A_s$). 
\begin{figure}
    \centering
    \includegraphics[width=0.7\linewidth]{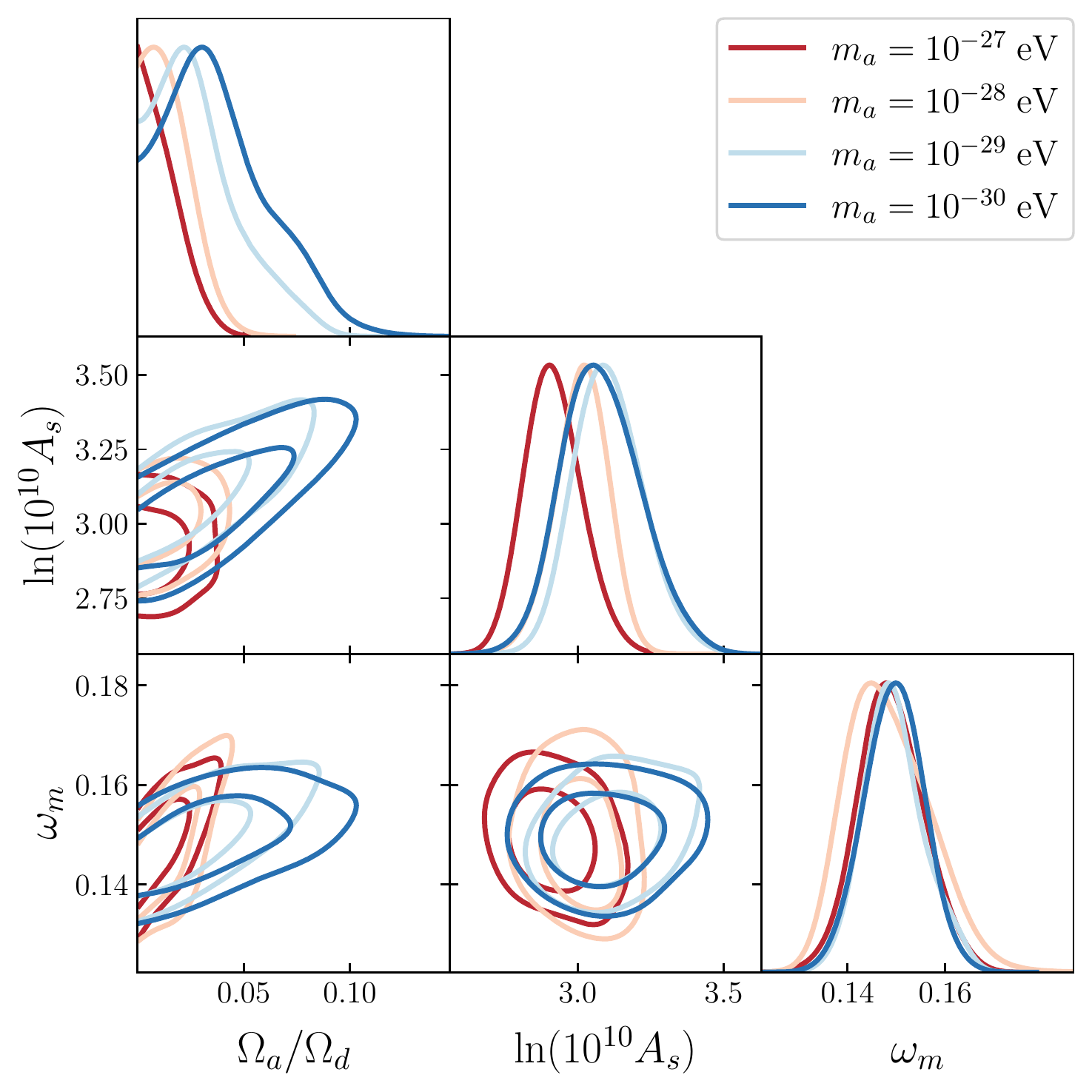}
    \caption{Joint posterior distribution for the power spectrum amplitude, the total matter density, and the axion density for four axion masses highlighting the dependence of the degeneracies on the particle mass.}
    \label{fig:As_OmM_degeneracy}
\end{figure}
This can be clearly seen in Fig.~\ref{fig:As_OmM_degeneracy} where we show the joint posterior distribution for the axion fraction and the power spectrum amplitude. We note that for a mass of $10^{-27}$ eV with $k_{J,\mathrm{eq}}(m_a)/k_\mathrm{min}\approx 1.7$, the parameters appear independent while in the case with a mass of $10^{-30}$ eV and $k_{m}(m_a)/k_\mathrm{min}<0.2$, we see a very strong degeneracy between the two parameters. Also in Fig.~\ref{fig:posterior_axions}, we observe a slight preference for a non-zero axion fraction in the marginalized posterior. This is caused by opening the $A_s$ degeneracy and marginalising over $\Omega_d$ and it was also observed in Ref.~\cite{Hlozek2018UsingMatter}. We note that the axion fraction for a $10^{-32}$ eV axion is not degenerate with either $\omega_m$ or $A_s$ which explains why a peak in the posterior distribution is not observed for that mass in Fig.~\ref{fig:posterior_axions}. 

\begin{figure}
    \centering
    \includegraphics[width=0.6\linewidth]{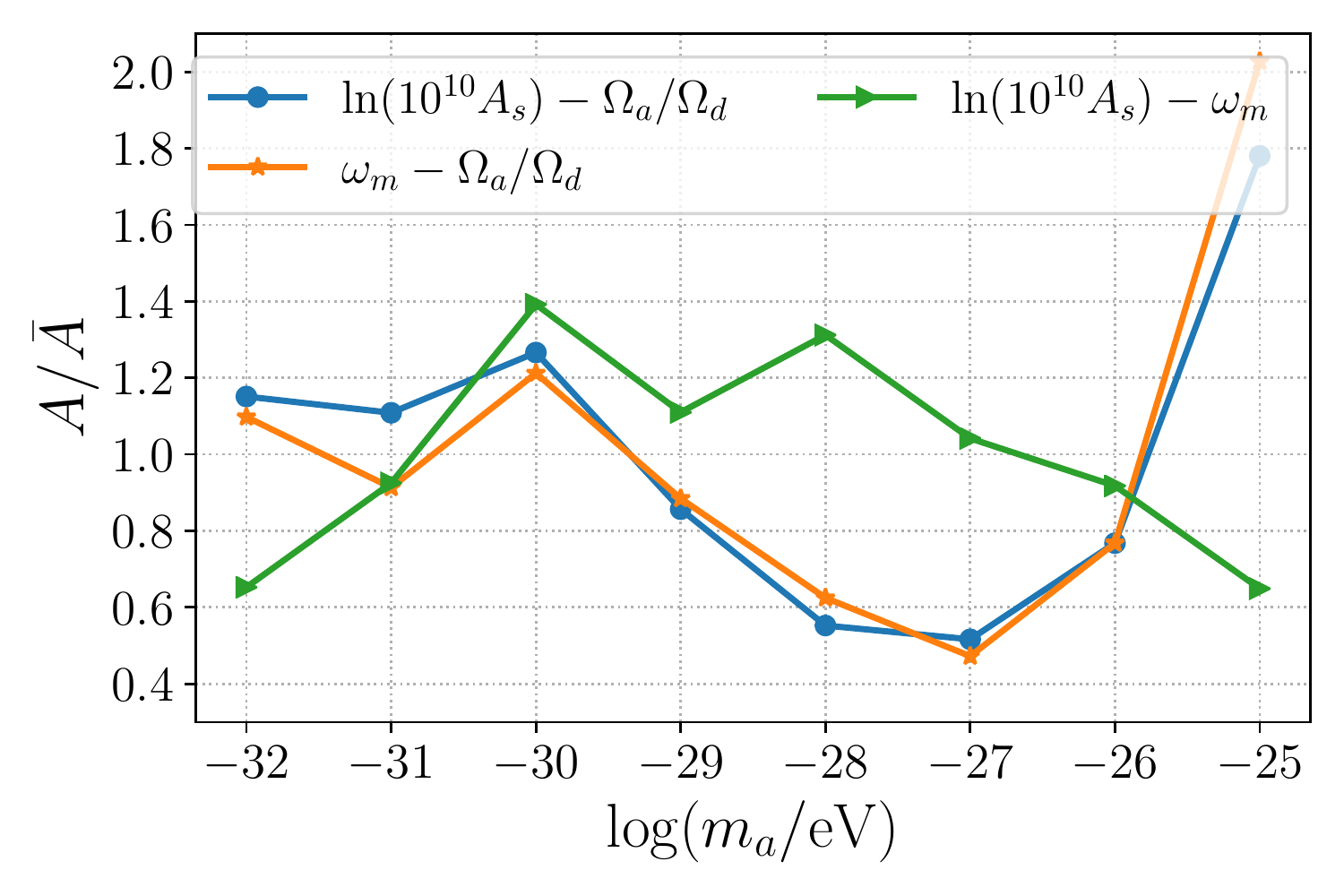}
    \caption{Area of joint posterior ellipses for three pairs of parameters. The areas are rescaled by the average taken over axion masses to allow for a better comparison. We note a simultaneous weakening of the constraints on $\Omega_a/\Omega_d,\;\ln(10^{10}A_s),\;\omega_m$ at a mass of $10^{-30}$ eV.}
    \label{fig:all_corr}
\end{figure}

Heavier axion masses ($10^{-27}-10^{-28}$ eV) also suffer from another degeneracy with the total matter density as shown in Fig.~\ref{fig:As_OmM_degeneracy}. This is similar to the degeneracy observed for massive neutrinos as the power spectrum below a characteristic scale is partially suppressed. In the case of axions, this suppression as a function of axion concentration and axion mass is shown in Fig.~\ref{fig:linear_pk} while for a single massive neutrino species, the suppression can be approximated by~\cite{Hu1998WeighingNeutrinos}
\begin{align}
    \frac{\Delta P(k)}{P(k)} \approx -0.08 \omega_m^{-1}\left(\frac{ m_\nu}{1\;\mathrm{eV}}\right)\;\;\;\mathrm{for}\;k>k_\mathrm{nr},\label{eq:neutrino_suppress}
\end{align}
where $m_\nu$ is the neutrino mass and where $k_\mathrm{nr}$ is the horizon scale at the time the neutrinos become non-relativistic which can be expressed as
\begin{align}
    k_\mathrm{nr} \approx 0.026 \omega_m^{1/2} \left(\frac{ m_\nu}{1\;\mathrm{eV}}\right)^{1/2}\;\mathrm{Mpc}^{-1}.
\end{align}
Although there exists no analytical approximation of the suppression level of the power spectrum due to axions there exists a few semi-analytic descriptions of  Eq.~(\ref{eq:neutrino_suppress}) which can be found in Ref.~\cite{Amendola2006DarkMatter,Marsh2010UltralightScalar,Marsh2012UltralightAxion}. The similarities between massive neutrinos and axions have been the subject of extensive studies using a collection of other cosmological probes~\cite{Marsh2012UltralightAxion}. We note that the axions' Jeans scale and the neutrinos $k_\mathrm{nr}$ are identical in their particle mass dependence. Also, we can observe from the BOSS data that axions seem to suffer from a similar relationship between the total matter density and the axion concentration where an increase in the matter density can (partially) compensate for the loss of clustering on small scales. This is evident in the case of neutrinos when taking the limit $\omega_m \gg m_\nu/\mathrm{eV}$ and for axions this can be read from Fig.~\ref{fig:As_OmM_degeneracy}. Finally, it is worth noting that neutrinos are also degenerate with the power spectrum amplitude when studying the Lyman-$\alpha$ forest~\cite{Pedersen2020MassiveNeutrinos} as are the lightest axions in our study. It is worth noting that allowing the sum of the neutrino masses to vary may also weaken some of the constraints in this work for some axion masses for which $k_m\sim k_\mathrm{nr}$.

Fig.~\ref{fig:all_corr} summarizes the degeneracies between the axion fraction and the cosmological parameters which have been varied in this study. In order to quantify the impact of varying a given parameter on our constraints on the axion fraction, we use the ellipse area given by~\cite{Coe2009FisherMatrices}
\begin{align}
    A(X,Y) = \pi \sigma_X \sigma_Y \sqrt{1-\rho(X,Y)^2}
\end{align}
where $X\in\left\{\Omega_a/\Omega_d,\ln(10^{10} A_s),\omega_m\right\}$, $\sigma_i$ is the 1$\sigma$ error on the parameter $i$, and where $\rho$ is the usual correlation coefficient
\begin{align}
    \rho(X,Y) = \frac{\sigma_{XY}}{\sigma_{X}\sigma_{Y}}.
\end{align}
This derived quantity captures degeneracies with axions but it also accounts for the capacity of the data to constrain the parameters correlated with axions. This is justified since a degeneracy with a very well constrained parameter will not lead to a wider posterior distribution for the axion fraction. We note a strong increase in area for axions with a mass of $m_a\sim10^{-30}$ eV for the ellipse taken from the joint posterior distribution of the matter density and the power spectrum amplitude. This indicates a degeneracy which loosens the constraints on the axion fraction at that mass. This can be seen in Fig.~\ref{fig:combined_constraint} as a wider permitted region at $10^{-30}$ eV. This explains why the one and two sigma contours are not monotonically decreasing from $10^{-32}$ eV as one would naively expect.

\begin{figure}
    \centering
    \includegraphics[width=0.9\linewidth]{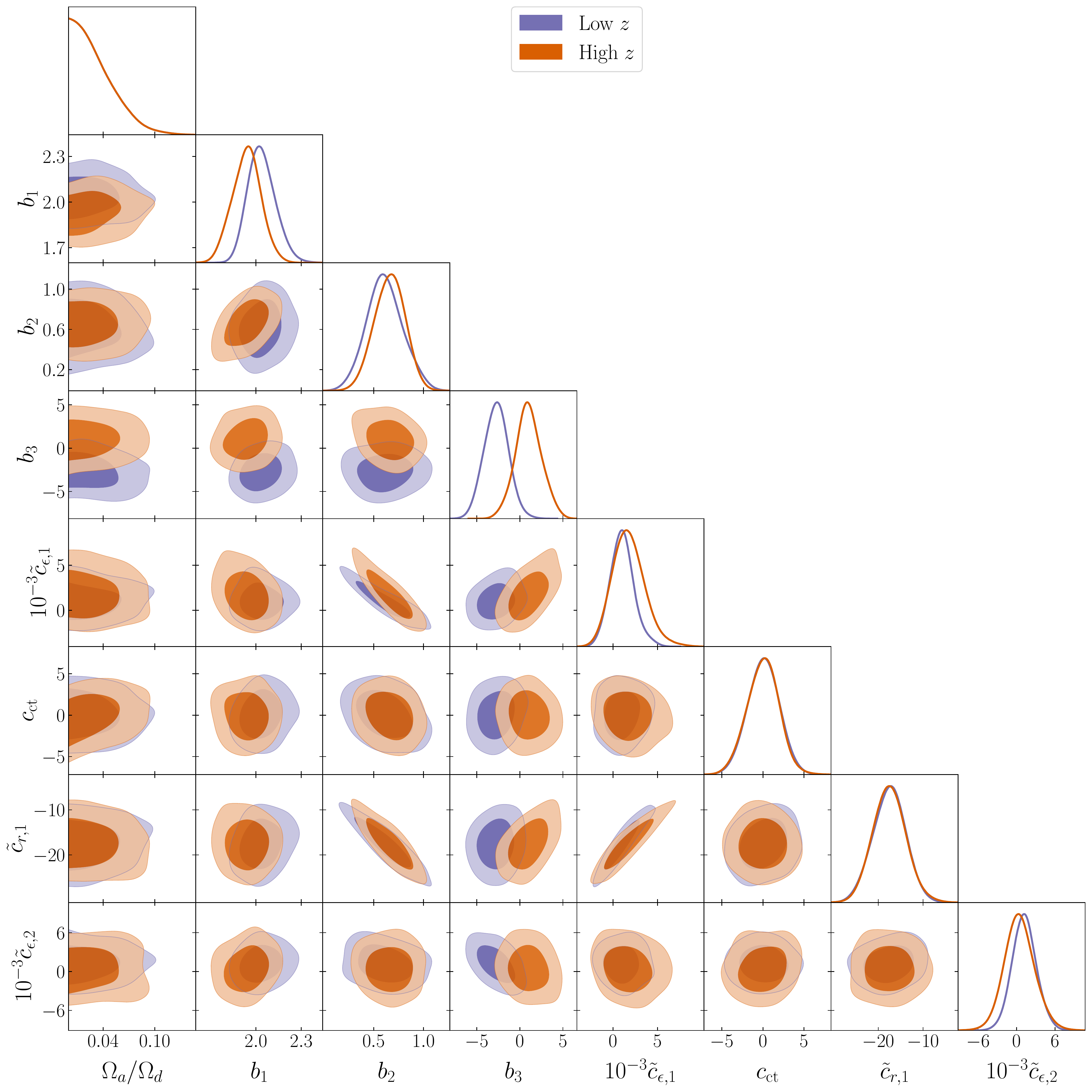}
    \caption{Marginalized 2D posterior distributions for the model's nuisance parameters (see Table~\ref{tab:priors}) and the axion fraction (leftmost column) for an axion mass of $10^{-27}$ eV. We note no severe degeneracies between the axion fraction and the nuisance parameters for either low or high redshifts. Similar results are obtained for all other axion masses in our analysis.}
    \label{fig:nuisance_degen}
\end{figure}

Finally, we look for degeneracies between the axion fraction and the nuisance parameters of the model. We show the marginalised posterior distribution of the nuisance parameters along with the axion fraction in Fig.~\ref{fig:nuisance_degen}. Given the shape of the galaxy monopoles in Fig.~\ref{fig:mon_quad_model}, we expected to find strong degeneracies between the axion fraction and the linear galaxy bias $b_1$. However, we do not observe such degeneracy for either of the redshift slices in the data. Another study based on a similar model finds no strong degeneracy between the galaxy biases and the power spectrum amplitude~\cite{Ivanov2020CosmologicalParameters} which is itself degenerate with the axion fraction for low axion masses ($\leq 10^{-29}$ eV). 
We conclude that our choice of model based on the EFTofLSS was well suited for probing ultralight axions using galaxy clustering despite the high number of nuisance parameters associated with working with such datasets.



\subsection{Galaxy Surveys with a CMB Prior}
We now recalculate the constraints obtained in the previous section by replacing the BBN prior on $\omega_b$ with a CMB prior on all cosmological parameters and the axion fraction. The CMB prior is obtained from the chains from the analysis in Ref.~\cite{Hlozek2018UsingMatter}. This work obtained constraints on axions using a combination of temperature, polarization and lensing data from the CMB. It also considered adiabatic initial conditions and repeated the analysis including axion-sourced isocurvature perturbations. In this work, we use the former as our prior since we assume adiabatic initial conditions for the density perturbations. 
The chains are once again run for each mass bin between $10^{-25}\;\mathrm{eV}$ and $10^{-32}\;\mathrm{eV}$. The resulting updated constraint plot is shown in Fig.~\ref{fig:combined_constraint}.

\begin{figure}
    \centering
    \includegraphics[width=0.4\linewidth]{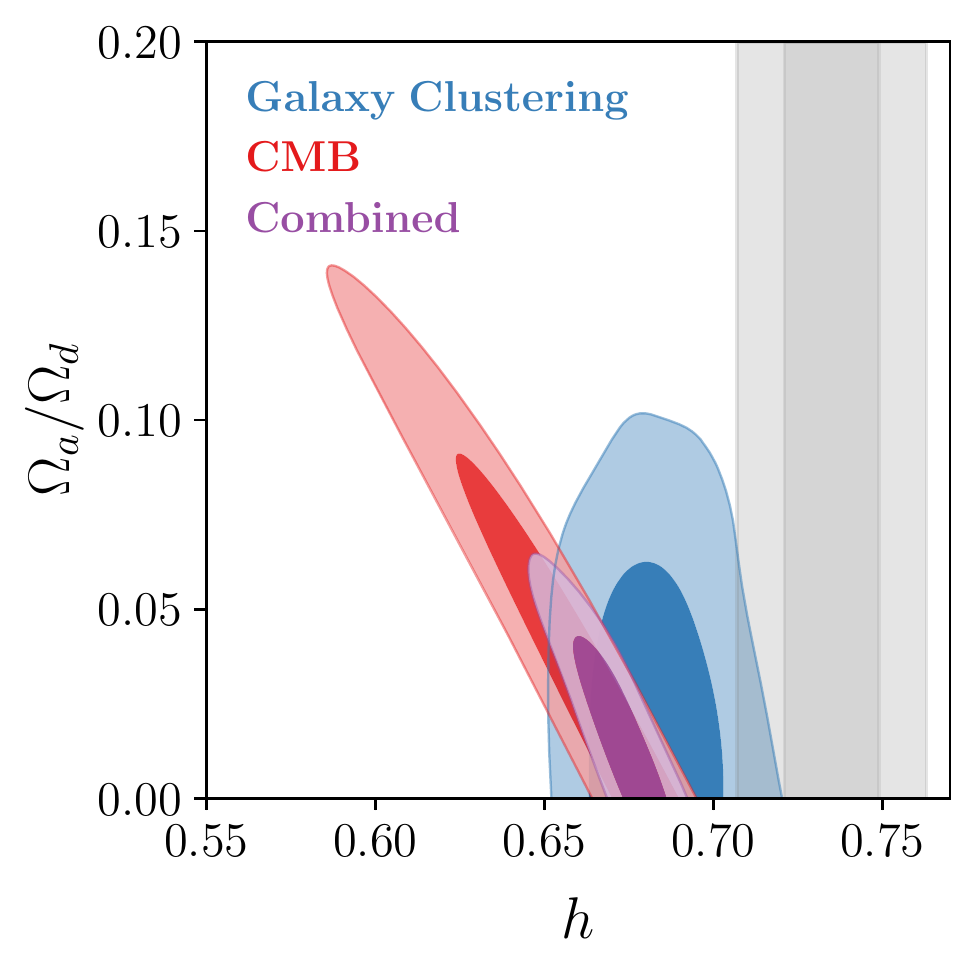}
    \caption{Joint posterior distributions for an axion with a mass of $10^{-32}$ eV for three experimental setups. We note an improvement on the constraint on the axion fraction when breaking the degeneracy with $H_0$ present with the CMB data. The gray shaded area represent the confidence interval for $h$ from the SH0ES measurement~\cite{Reid2019AnImproved}.}
    \label{fig:combined_triangle}
\end{figure}
\begin{figure}
    \centering
    \includegraphics[width=.8\linewidth]{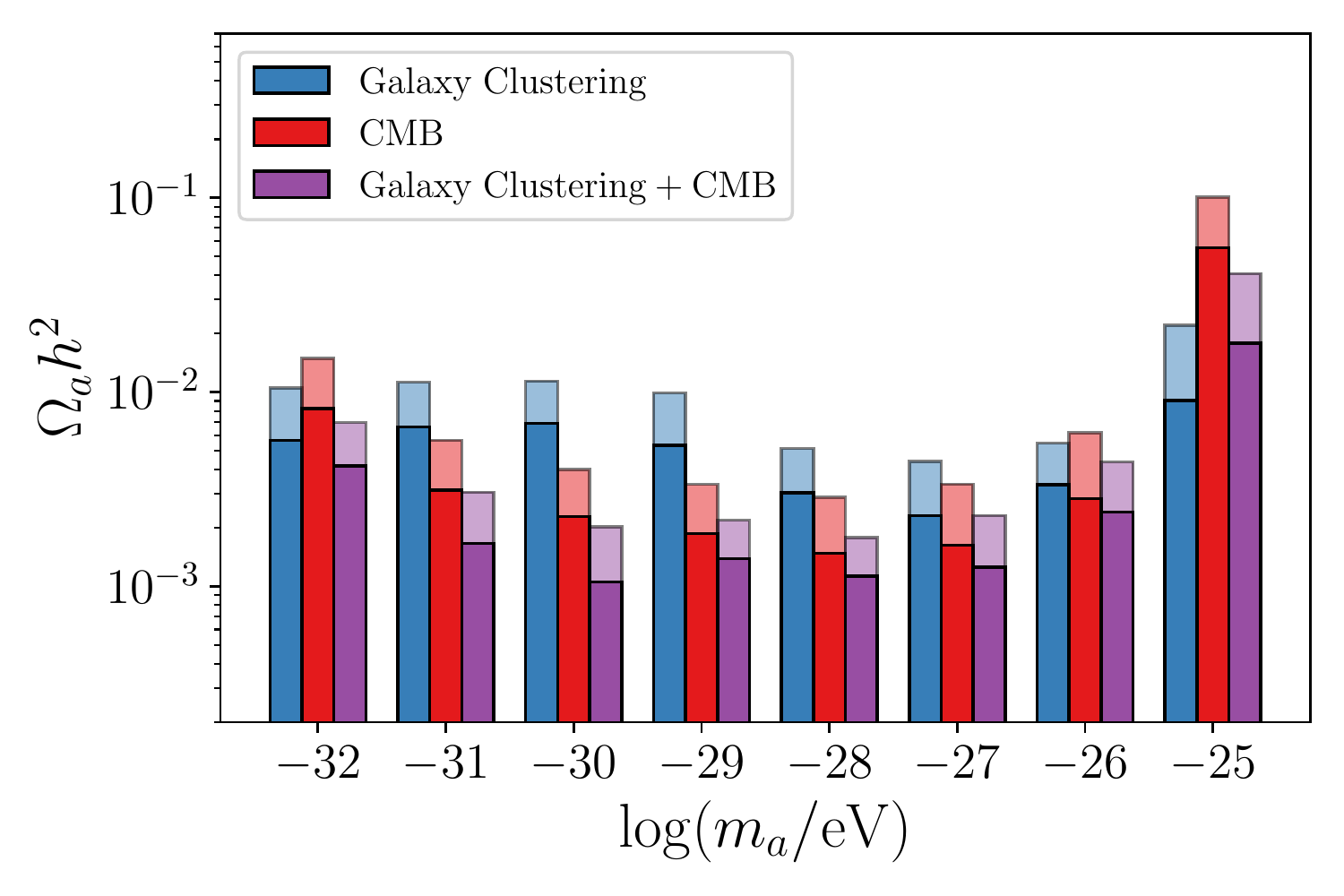}
    \caption{68\% (dark-colored) and 95\% (light-colored) confidence level bounds on the axion density from the CMB data, galaxy clustering and the combined measurements.}
    \label{fig:omega_a_bar_plot}
\end{figure}

\begin{table}[H]
    \centering
    \begin{tabular}{ |cccc| } 
        \hline
        Mass [$\log(m_a/\mathrm{eV})$] & CMB (Planck 2015) & Galaxy Clustering & Combined Measurements \\
        \hline
        $-25$ & $<0.101$ & $<0.022$ & $<0.041$ \\
        $-26$ & $<0.006$ & $<0.005$ & $<0.004$ \\
        $-27$ & $<0.003$ & $<0.004$ & $<0.002$ \\
        $-28$ & $<0.003$ & $<0.005$ & $<0.002$ \\
        $-29$ & $<0.003$ & $<0.010$ & $<0.002$ \\
        $-30$ & $<0.004$ & $<0.011$ & $<0.002$ \\
        $-31$ & $<0.006$ & $<0.011$ & $<0.003$ \\
        $-32$ & $<0.015$ & $<0.011$ & $<0.007$ \\
        \hline
    \end{tabular}
    \caption{\label{tab:results} Numerical values of the 95\% C.L. upper bounds on $\Omega_ah^2$ from three different experimental configurations as shown in Fig.~\ref{fig:omega_a_bar_plot}.}
    \label{tab:2_sigma_bounds}
\end{table}

The joint analysis allows for much tighter constraints on the axion fraction by breaking degeneracies between $\Omega_a$ and other cosmological parameters. See Fig.~\ref{fig:combined_triangle} for an example of broken degeneracy in the case of $H_0$ with $10^{-32}$ eV axions. When considering only galaxy clustering data, the presence of axions increases the error bars on $H_0$ thus initially alleviating the tension with late time measurements. This is very similar to the impact of the inclusion of a higher effective number of relativistic species of Ref.~\cite{Bernal2016TheTrouble}. When including the CMB prior however, we observe that a large fraction of axions at a mass of $10^{-32}$ eV leads to an even higher discrepancy with SN1a measurements of $H_0$. This contrasts with other ultralight scalars with modified field potentials which have been proposed as solutions to this tension~\cite{Poulin2018CosmologicalImplications}. There is one case however where the addition of a CMB prior does not lead to stronger constraints on the axion fraction.
\begin{figure}
    \centering
    \includegraphics[width=0.45\linewidth]{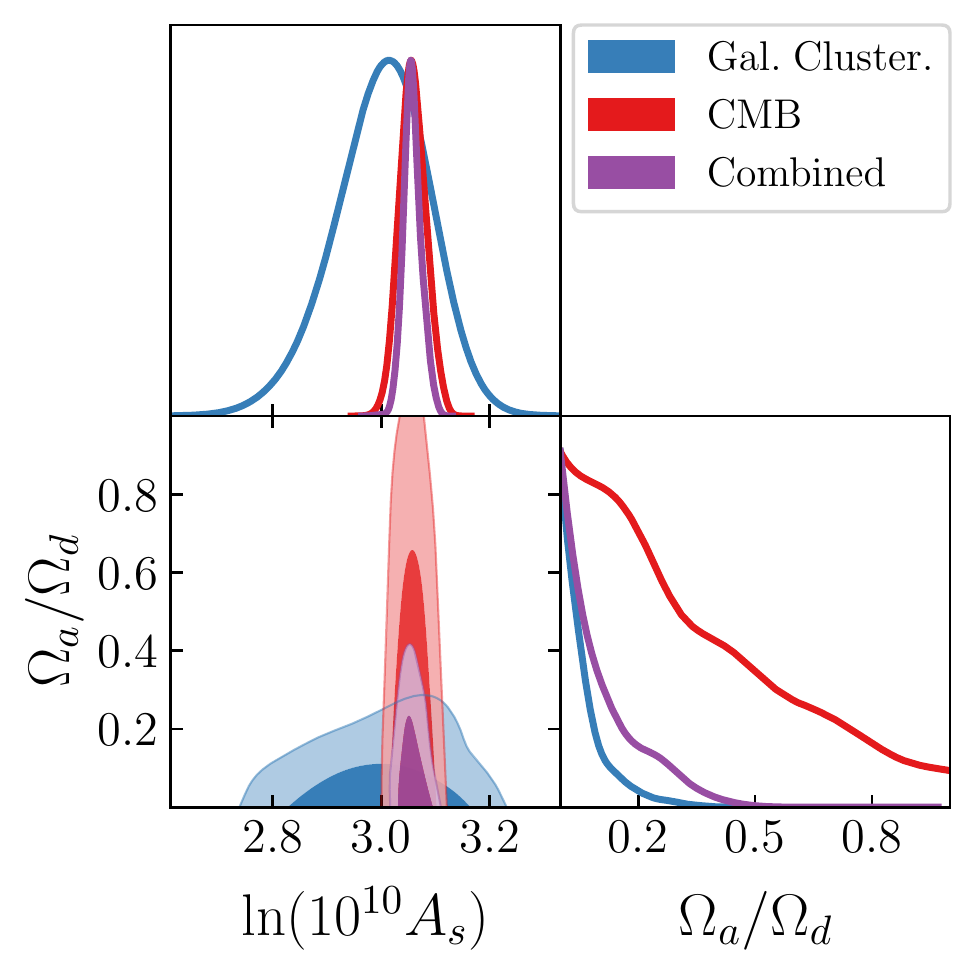}
    \caption{Degeneracy between the power spectrum amplitude and the axion fraction for an axion mass of $10^{-25}$ eV. We observe that the CMB prior favours a region of parameter space for which the galaxy clustering data has a wider uncertainty on the axion fraction.}
    \label{fig:1_25_degen}
\end{figure}
Indeed, the constraint on the axion fraction with the mass of $10^{25}$ eV worsens when adding a CMB prior. This is due to the fact that the CMB prior favours a higher value of $A_s$ which is slightly degenerate with the axion fraction at that mass as shown in Fig.~\ref{fig:1_25_degen}. Another contributing factor is that the CMB prior does not constrain the axion fraction as well as for the axion masses below $10^{-25}$ eV. Performing a joint likelihood analysis rather than imposing a prior on the cosmological parameters may allow for stronger constraints for this mass bin and is left for future work. We note however that galaxy clustering measurements alone improve existing constraints on the axion fraction at that mass by over a factor of 4.5 (see Table~\ref{tab:2_sigma_bounds}).

Combining CMB data with galaxy clustering allows us to overcome the degeneracies present in the CMB analysis. The improvements on the bounds on the axion density are displayed in Fig.~\ref{fig:omega_a_bar_plot}. There we see a factor of two improvement on CMB alone for most of the axion masses considered. A joint analysis of CMB and galaxy survey data was conducted in Ref~\cite{Hlozek2015AData} with the Planck and the WiggleZ~\cite{Blake2010TheWiggleZ} surveys and gave $\Omega_a/\Omega_d\leq 0.048$ for a similar mass range. This work was based on the 2013 data release and considered adiabatic initial conditions. Two factors contribute to improved constraints in the present work. The first is that the WiggleZ data were limited to the galaxy power spectrum while the BOSS data used includes both the  monopoles and quadrupoles. This allows us to detect the axions' anisotropic effects on galaxies and amplifies the size of our galaxy clustering dataset by about a factor of four. The second factor is the development of modeling techniques allowing us to model the power spectrum multipoles up to a scale of $k=0.23\;h/$Mpc while the initial work of Ref.~\cite{Hlozek2015AData} was limited to $k_\mathrm{max}=0.2\;h/$Mpc. This difference is most notable on the heaviest mass of $10^{-25}$ eV where our 95\% C.L. upper bound is improved by a factor of $\sim 1.6$ since the power spectrum cut-off manifests on the smallest scales.

\section{Discussion}\label{sec:discussion}
We introduced ultralight axion physics in a galaxy clustering model by modifying the linear power spectrum and then calculating higher order corrections using the EFTofLSS approach through the use of the public code \texttt{PyBird}. To increase calculation efficiency, we also developed a way to obtain the linear matter power spectrum in the presence of ultralight axions using interpolations thus significantly decreasing the total computation time. We first ran the canonical tests for galaxy modeling using the MultiDark-Patchy mocks simulation suite and verified that our model was able to recover $\Lambda$CDM parameters in the absence of axions. We also generated our own mock catalogs using an HOD approach and the Lagrangian based simulation code \texttt{Peak-Patch}. This allowed us to assert that our model is sensitive to even small axion fractions of order $\Omega_a/\Omega_d\sim 0.05$. Results from simulations including $10^{-28}$ eV axions also showed that they could bias our estimates of $S_8$ if they are present but unaccounted for in our model. In the present work, we have not considered interactions between axions and the standard model. A typical interaction takes the form $\phi {\bf E}\cdot {\bf B}$. Such an interaction affects the polarisation of the CMB, and for $10^{-33}\text{ eV}\leq m_a\leq 10^{-28}\text{ eV}$ could explain the recently observed \emph{birefringence} in CMB anisotropies~\cite{Minami2020NewExtraction}. If this measurement is confirmed, this motivates a drive to improve searches for axions in this mass window using structure formation, which could allow an independent measurement of $m_a$ and $f_a$ impossible with birefringence alone. We note that birefringence may also be induced by a network cosmic strings if the Peccei-Quinn symmetry of the axion field is broken after inflation~\cite{Agrawal2020ACMB,Jain2021CMBBirefringence}.

Having tested our model in various axion and axion-free cosmologies, we ran multiple studies on the BOSS galaxy clustering dataset in redshift space. We performed 16 different runs using eight logarithmically spaced mass bins and two choices of priors. Our axion mass range extended over seven orders of magnitude from $10^{-32}$ eV to $10^{-25}$ eV. Axions lighter than this would behave as a dark energy component in the relevant regime. Axions heavier than the upper bound would have a Jeans scale much smaller than the semi-linear scales to which we are sensitive making them identical to CDM for $k\leq k_\mathrm{max}$. Constraining axions heavier than this mass scale would require a fully non-linear treatment which has not yet been fully explored in the mixed axion dark matter case. See Ref.~\cite{Lague2020EvolvingUltralight} for a semi-linear treatment at high redshift for initial conditions and Ref.~\cite{Schwabe2020SimulatingMixed} for the spherical collapse in mixed dark matter case. Using a BBN prior on the baryon density only, we arrived at constraints matching that of the CMB analysis for the lighter masses, but improved them significantly for masses above $10^{-29}$ eV. Using a CMB prior on all cosmological parameters including the axion parameters, we arrived at the strongest constraint on the axion density with $\Omega_a h^2 <0.022$ for the heaviest axions in this study. We also arrived at the bounds $\Omega_a h^2< 0.002$ for the mass window $10^{-30}\;\mathrm{eV} \leq m_a \leq 10^{-27}\;\mathrm{eV}$ and $\Omega_a h^2< 0.007$ for axions with $m_a=10^{-32}\;\mathrm{eV}$. These results are expressed in terms of the axion fraction in Fig.~\ref{fig:combined_constraint} and in terms of the axion relic density in Table~\ref{tab:2_sigma_bounds} and Fig.~\ref{fig:omega_a_bar_plot}. Combining CMB and large-scale structure data, we were able to break important degeneracies between axions and $\Lambda$CDM parameters and we arrived at the most stringent bounds on the axion density for this mass range to date. The 95\% C.L. bounds on the axion fraction obtained in this work approach those projected for the CMB-S4 observatory~\cite{Hlozek2017FutureCMB} alone but were attained using existing data thus showcasing the power of multi-probe approaches in cosmology.

\acknowledgments
We would like to thank Mona Dentler, Colin Hill, Evan McDonough, and Mikhail Ivanov for useful comments and suggestions. RH is a CIFAR Azrieli Global Scholar, Gravity \& the Extreme Universe Program, 2019, and a 2020 Alfred. P. Sloan Research Fellow. RH is supported by Natural Sciences and Engineering Research Council of Canada. RB is a CIFAR Fellow. AL, RB and RH are supported by Natural Sciences and Engineering Research Council of Canada. The Dunlap Institute is funded through an endowment established by the David Dunlap family and the University of Toronto. We acknowledge that the land on which the University of Toronto is built is the traditional territory of the Haudenosaunee, and most recently, the territory of the Mississaugas of the New Credit First Nation. We are grateful to have the opportunity to work in the community, on this territory. Computations were performed on the Niagara supercomputer~\cite{Loken2010LessonsLearned,Ponce2019DeployingA} at the SciNet HPC Consortium. SciNet is funded by: the Canada Foundation for Innovation; the Government of Ontario; Ontario Research Fund - Research Excellence; and the University of Toronto. The data visualization in this work used the software package \texttt{getdist}~\cite{Lewis2019GetDist}. The MCMC chains were run in parallel with MPI through the use of the \texttt{schwimmbad} library~\cite{Price-Whelan2017schwimmbad}.

\appendix
\section{More on Perturbation Theory} \label{app:perturbations}
\subsection{Higher Order Terms with CDM Dynamics}
The first loop correction has two contributions and takes the form of~\cite{Bernardeau2002Large-Scale}
\begin{align}
    P_{1-\mathrm{loop}}(k) = P_{22}(k) + P_{13}(k),
\end{align}
where
\begin{align}
    P_{22}(k) &\equiv 2\int \frac{d^3q}{(2\pi)^3}\left[F_2(\mathbf{q},\mathbf{k}-\mathbf{q})\right]^2 P_\mathrm{lin}(|\mathbf{k}-\mathbf{q}|) P_\mathrm{lin}(q),\\
    P_{13}(k) &\equiv 6\int \frac{d^3q}{(2\pi)^3} F_3(\mathbf{q},-\mathbf{q},\mathbf{k}) P_\mathrm{lin}(k) P_\mathrm{lin}(q),
\end{align}
where $P_\mathrm{lin}(k)$ is the linear matter power spectrum\footnote{The linear matter power spectrum is often denoted as $P_{11}(k)$.} and where the functions $F_{2,3}$ are standard Eulerian perturbation theory kernels which can be found by recursion relations~\cite{Bernardeau2002Large-Scale}. To model the data, we must also consider the effects redshift space distortions and the fact that galaxies are biased tracers. We can model redshift space distortions by mapping the redshift space positions $\mathbf{x}_r$ from the real-space positions following
\begin{align}
    \mathbf{x}_r = \mathbf{x} + \frac{\hat{\mathbf{z}} \cdot \hat{\mathbf{v}}}{\mathcal{H}}\hat{\mathbf{z}},
\end{align}
where $\mathbf{v}$ is the tracer's velocity. The impact of redshift space distortions on biased tracers has been studied in detail in Ref.~\cite{Senatore2014RedshiftSpace}. The relationship between the halo overdensities in redshift space are then obtained from the real space results by
\begin{align}
    \delta_{h,r}(\mathbf{k}) = \delta_{h}(\mathbf{k}) + \int d^3x e^{-i\mathbf{k}\cdot\mathbf{x}} \left[\exp\left(-i\frac{k_z}{\mathcal{H}}v_{h,z}\right)-1\right] (1+\delta_h(\mathbf{x})).
\end{align}
Expanding the halo overdensity up to third order, we get
\begin{align}
    \delta_{h,r} = \delta_{h,r}^{(1)} + \delta_{h,r}^{(2)} + \delta_{h,r}^{(3)} + \delta^{(3)}_{h,r,\mathrm{counter}} +\delta^{(3)}_{h,r,\mathrm{stoch}} + ... .
\end{align}
where the two last terms account for the counter and stochastic terms arising from the renormalization of the power spectrum (need more details).
The halo power spectrum in redshift space is computed with loop-order corrections from the linear matter power spectrum by the inclusion of higher order terms and with the addition of counter and stochastic corrections. The full expression from Ref.~\cite{Perko2016BiasedTracers} reads
\begin{align}
\left\langle\delta_{h, r}(\mathbf{k}) \delta_{h, r}(\mathbf{k})\right\rangle&=\left\langle\delta_{h, r}^{(1)} \delta_{h, r}^{(1)}\right\rangle+\left\langle\delta_{h, r}^{(2)} \delta_{h, r}^{(2)}\right\rangle+2\left\langle\delta_{h, r}^{(1)} \delta_{h, r}^{(3)}\right\rangle+\left\langle\delta_{h, r} \delta_{h, r}\right\rangle_{\mathrm{ct}}+\left\langle\delta_{h, r} \delta_{h, r}\right\rangle_{\epsilon} \\
&= b_1^{2} P_\mathrm{lin}(k)+2 \int d^{3} q \left[K_{h, r}^{(2)}(b_1,b_2,b_4;\mathbf{q}, \mathbf{k}-\mathbf{q})_\mathrm{sym}\right]^{2} P_\mathrm{lin}(|\mathbf{k}-\mathbf{q}|) P_\mathrm{lin}(q) \nonumber \\&+6 b_1 \int d^{3} q K_{h, r}^{(3)}(b_1,b_3;\mathbf{q},-\mathbf{q}, \mathbf{k})_\mathrm{sym} P_\mathrm{lin}(q) P_\mathrm{lin}(k)+\left\langle\delta_{h, r} \delta_{h, r}\right\rangle_{\mathrm{ct}}+\left\langle\delta_{h, r} \delta_{h, r}\right\rangle_{\epsilon},
\end{align}
where $b_{1,2,3,4}$ are the bias parameters which will be varied in our analysis and $K_{h,r}^{(2,3)}$ are the symmetrized versions of the redshift-space halo kernels whose complete derivation can be found in Ref.~\cite{Perko2016BiasedTracers}. The redshift-space kernels are
\begin{align}
\begin{split}
K_{h, r}^{(2)}\left(\mathbf{q}_{1}, \mathbf{q}_{2}\right)=K_{\delta_{h}}^{(2)} &\left(\mathbf{q}_{1}, \mathbf{q}_{2}\right)+f \mu^{2} K_{\theta_{h}}^{(2)}\left(\mathbf{q}_{1}, \mathbf{q}_{2}\right) \\
&+\frac{1}{2} b_1 \mu f\left(\frac{k q_{2 z}}{q_{2}^{2}}+\frac{k q_{1 z}}{q_{1}^{2}}\right) K_{\theta_{h}}^{(1)}\left(\mathbf{q}_{1}\right) +\frac{1}{2} \mu^{2} f^{2} \frac{k^{2} q_{1 z} q_{2 z}}{q_{1}^{2} q_{2}^{2}} K_{\theta_{h}}^{(1)}\left(\mathbf{q}_{1}\right) K_{\theta_{h}}^{(1)}\left(\mathbf{q}_{2}\right)
\end{split}
\end{align}
\begin{align}
\begin{split}
K_{h, r}^{(3)}\left(\mathbf{q}_{1}, \mathbf{q}_{2}, \mathbf{q}_{3}\right)=& K_{\delta_{h}}^{(3)}\left(\mathbf{q}_{1}, \mathbf{q}_{2}, \mathbf{q}_{3}\right)+f \mu^{2} K_{\theta_{h}}^{(3)}\left(\mathbf{q}_{1}, \mathbf{q}_{2}, \mathbf{q}_{3}\right) \\
&+b_1\mu f\left(\frac{k q_{3 z}}{q_{3}^{2}}\right) K_{\delta_{h}}^{(2)}\left(\mathbf{q}_{1}, \mathbf{q}_{2}\right) K_{\theta_{h}}^{(1)}\left(\mathbf{q}_{3}\right)+\mu f\left(\frac{k\left(q_{1 z}+q_{2 z}\right)}{\left(\mathbf{q}_{1}+\mathbf{q}_{2}\right)^{2}}\right) K_{\theta_{h}}^{(2)}\left(\mathbf{q}_{1}, \mathbf{q}_{2}\right) \\
&+\frac{1}{2} \mu^{2} f^{2}\left(\frac{k q_{1 z}}{q_{1}^{2}} \frac{k q_{2 z}}{q_{2}^{2}}\right) K_{\theta_{h}}^{(1)}\left(\mathbf{q}_{1}\right) K_{\theta_{h}}^{(1)}\left(\mathbf{q}_{2}\right) K_{\delta_{h}}^{(1)}\left(\mathbf{q}_{3}\right) \\
&+\mu^{2} f^{2}\left(\frac{k\left(q_{1 z}+q_{2 z}\right)}{\left(\mathbf{q}_{1}+\mathbf{q}_{2}\right)^{2}} \frac{k q_{3 z}}{q_{3}^{2}}\right) K_{\theta_{h}}^{(2)}\left(\mathbf{q}_{1}, \mathbf{q}_{2}\right) K_{\theta_{h}}^{(1)}\left(\mathbf{q}_{3}\right) \\
&+\frac{1}{6} \mu^{3} f^{3}\left(\frac{k q_{1 z}}{q_{1}^{2}} \frac{k q_{2 z}}{q_{2}^{2}} \frac{k q_{3 z}}{q_{3}^{2}}\right) K_{\theta_{h}}^{(1)}\left(\mathbf{q}_{1}\right) K_{\theta_{h}}^{(1)}\left(\mathbf{q}_{2}\right) K_{\theta_{h}}^{(1)}\left(\mathbf{q}_{3}\right),
\end{split}
\end{align}
where the dependence on the bias parameters is mae explicit in the higher order real-space kernels. For the halo overdensity $\delta_h$ they are given by
\begin{align}
\begin{split}
K_{\delta_{h}}^{(2)}(k, q, x)_{\mathrm{sym}}=& \frac{b_1-2 q^{3}+k^{3} x+4 k q^{2} x-k^{2} q\left(1+2 x^{2}\right)}{k^{2}+q^{2}-2 k q x} \\
&+\frac{b_2}{7} \frac{7 q^{2}-14 k q x+k^{2}\left(5+2 x^{2}\right)}{k^{2}+q^{2}-2 k q x}+b_4
\end{split}
\end{align}
\begin{align}
\begin{split}
K_{\delta_{h}}^{(3)}(k, q)_{\mathrm{UV}-\mathrm{sub}, \mathrm{sym}}=\frac{b_1}{504 k^{3} q^{3}}\left(-38 k^{5} q+48 k^{3} q^{3}-18 k q^{5}+9\left(k^{2}-q^{2}\right)^{3} \log \left[\frac{k-q}{k+q}\right]\right) \\
+\frac{b_3}{756 k^{3} q^{5}}\left(2 k q\left(k^{2}+q^{2}\right)\left(3 k^{4}-14 k^{2} q^{2}+3 q^{4}\right)+3\left(k^{2}-q^{2}\right)^{4} \log \left[\frac{k-q}{k+q}\right]\right).
\end{split}
\end{align}
And for the halos' velocity divergence $\theta_h$, we have
\begin{align}
K_{\theta_{h}}^{(1)}(k, q, x)_{\mathrm{sym}} &=1, \\
K_{\theta_{h}}^{(2)}(k, q, x)_{\mathrm{sym}} &=\frac{k^{2}\left(7 k x-q\left(1+6 x^{2}\right)\right)}{14 q\left(k^{2}+q^{2}-2 k q x\right)}, \\
K_{\theta_{h}}^{(3)}(k, q, x)_{\mathrm{UV}-\mathrm{sub}, \mathrm{sym}} &=\frac{12 k^{7} q-82 k^{5} q^{3}+4 k^{3} q^{5}-6 k q^{7}+3\left(k^{2}-q^{2}\right)^{3}\left(2 k^{2}+q^{2}\right) \log \left[\frac{k-q}{k+q}\right]}{504 k^{3} q^{5}},
\end{align}
where UV-sub denotes the term with the high $k$ modes subtracted. We once again refer readers to Ref.~\cite{Perko2016BiasedTracers} for more detail.
The counterterm contribution is given by
\begin{align}
    \left\langle\delta_{h, r}(\mathbf{k}) \delta_{h, r}(\mathbf{k})\right\rangle_{\mathrm{ct}} \nonumber &= 2 P_\mathrm{lin}(k)\left(b_{1}+f \mu^{2}\right)\\&\;\;\;\times\left(\mu^{2}\left(\frac{k}{k_{\mathrm{M}}}\right)^{2} \tilde{c}_{r, 1}+\mu^{4}\left(\frac{k}{k_{\mathrm{M}}}\right)^{2} \tilde{c}_{r, 2}+c_{\mathrm{ct}}\left(\frac{k}{k_{\mathrm{NL}}}\right)^{2}\right),
\end{align}
where $f\equiv d\ln D/d\ln a$ is the logarithmic derivative of the linear growth factor $P(k,t) = \frac{D^2(t)}{D^2(t_0)}P(k,t_0)$, where $k_m$ is the scale of halos (more details here), and where $k_\mathrm{NL}$ is the scale at which non-linearities begin to dominate. The cosine of the angle between the line-of-sight and and the wavevector is denoted $\mu$ and $\{\tilde{c}_{r, 1}, \tilde{c}_{r, 2}, c_{\mathrm{ct}} \}$ is the set of counterterms which will be varied in our analysis.
\begin{align}
    \left\langle\delta_{h, r} \delta_{h, r}\right\rangle_{\epsilon} = \frac{1}{\bar{n}_W}\left[c_{\epsilon,1} + (c_{\epsilon,2}+c_{\epsilon,2}f\mu^2)\left(\frac{k}{k_\mathrm{M}}\right)^2\right],
\end{align}
where $\bar{n}_W$ is the typical halo density which we approximate using~\cite{Baldauf2016OnTheReach}
\begin{align}
    \bar{n}_W = \left(\int dM \frac{M^2}{\bar{\rho}^2}\frac{dn}{dM}\right)^{-1} \approx 330\;h^3/\mathrm{Mpc}^3.
\end{align}
Given that $k_\mathrm{M}\approx 0.7\; h/\mathrm{Mpc}$ and $k_\mathrm{max}\leq 0.23\; h/\mathrm{Mpc}$, we find little scale dependence on the stochastic terms and replace them with a single constant shot noise contribution to the monopole term along with a quadratic high-momentum term and absorb the contribution of the inverse halo density giving
\begin{align}
    \left\langle\delta_{h, r} \delta_{h, r}\right\rangle_{\epsilon} \approx \tilde{c}_{\epsilon,1} + \tilde{c}_{\epsilon,2}  \left(\frac{k}{k_{\mathrm{M}}}\right)^{2}.
\end{align}
And this completes our model. We thus have initially nine free parameters aside from the axion and cosmological parameters which are $\{b_{1,2,3,4}, c_\mathrm{ct}, \tilde{c}_{r,1}, \tilde{c}_{r,2}, \tilde{c}_{\epsilon,1}, \tilde{c}_{\epsilon,2}\}$. We identify which of the parameters we fix and the priors used for each of the varied ones in Section~\ref{sec:methods}.

\subsection{Wave Effects on Loop Corrections}
In this section, we tackle the importance of the dynamical effects of ultralight axions. Due to their quantum diffusion, axions do not cluster in the same way as collisionless dark matter and care must be taken if we wish to use the EFTofLSS which was designed for CDM. The wave effects on the clustering of axions is encoded at first order in the linear growth factor which is found by solving~\cite{Marsh2016AxionCosmology}
\begin{align}
    \ddot{D}(k,a) + 2H\dot{D}(k,a) +\left(\frac{\hbar^2k^4}{4m_a^2a^2}-4\pi G \bar{\rho}\right)D(k,a) = 0,
\end{align}
where the overdot denotes a derivative with respect to time. For simplicity, we will assume an Einstein deSitter cosmology from now on. We note that we recover the scale-independent CDM growth factor $D_\mathrm{CDM}(a)\propto a$ as we go to large scales. On small scales, the growth of perturbations is suppressed and rapidly oscillates around zero. Therefore, the assessment the importance of wave effects will have to be conducted for each scale. Integrating the differential equation for the growth factor from an initial $a_\mathrm{in}$, we have the solution~\cite{Chavanis2012GrowthMatter, Suarez2015HydrodynamicAnalysis,Li2019NumericalModel}
\begin{align}
    D(k,a) = \left(\frac{a_\mathrm{in}}{a}\right)^{1/4} \frac{J_{-5/2}\left(\hbar k^2/m_a H_0\sqrt{a}\right)}{J_{-5/2}\left(\hbar k^2/m_a H_0\sqrt{a_\mathrm{in}}\right)},
\end{align}
where the Bessel function of the first kind is given by
\begin{align}
J_{-5/2}(x) = \sqrt{\frac{2}{\pi x}}\bigg(\frac{3\cos x}{x^2}+\frac{3\sin x}{x}-\cos x\bigg).
\end{align}
Taking the $k\to 0$ limit, we get a constant growth equal to $a/a_\mathrm{in}$ as expected. Deviations from the CDM growth on small scales is what alters the kernels of in the integrals for the 1-loop corrections for the matter power spectrum. However, these deviations only affect the part of the matter which is composed of axions and it was shown earlier that the axion perturbations at the linear level are suppressed below their Jeans scale at matter-radiation equality. We argue here that \emph{the loss of power in the perturbations on small scales erases the wave effects before they become dominant.} Let us define the quantum corrections to the growth factor as
\begin{align}
    D_Q \equiv D(k,a) - D_\mathrm{CDM}(a).
\end{align}
The final amplitude of total quantum corrections depend on the size of the axion perturbations which they affect. Therefore, the amplitude of the quantum corrections at linear order are proportional to the product of the growth factor and the perturbations at each scale. If we define the relative corrections normalized with the CDM perturbations we have
\begin{align}
    \mathcal{Q}^{(1)}(k,a) = \left(\frac{D_Q(k,a)}{D_\mathrm{CDM}(a)}\right)^2\frac{\left\langle\delta^{(1)}_a\delta^{(1)}_a\right\rangle}{\left\langle\delta^{(1)}_c\delta^{(1)}_c\right\rangle},
\end{align}
where $\delta_{a,c}^{(i)}$ is the $i^\mathrm{th}$ order perturbations and where $a$ and $c$ denote the axion and CDM components respectively. 
\begin{figure}
    \centering
    \includegraphics[width=0.65\linewidth]{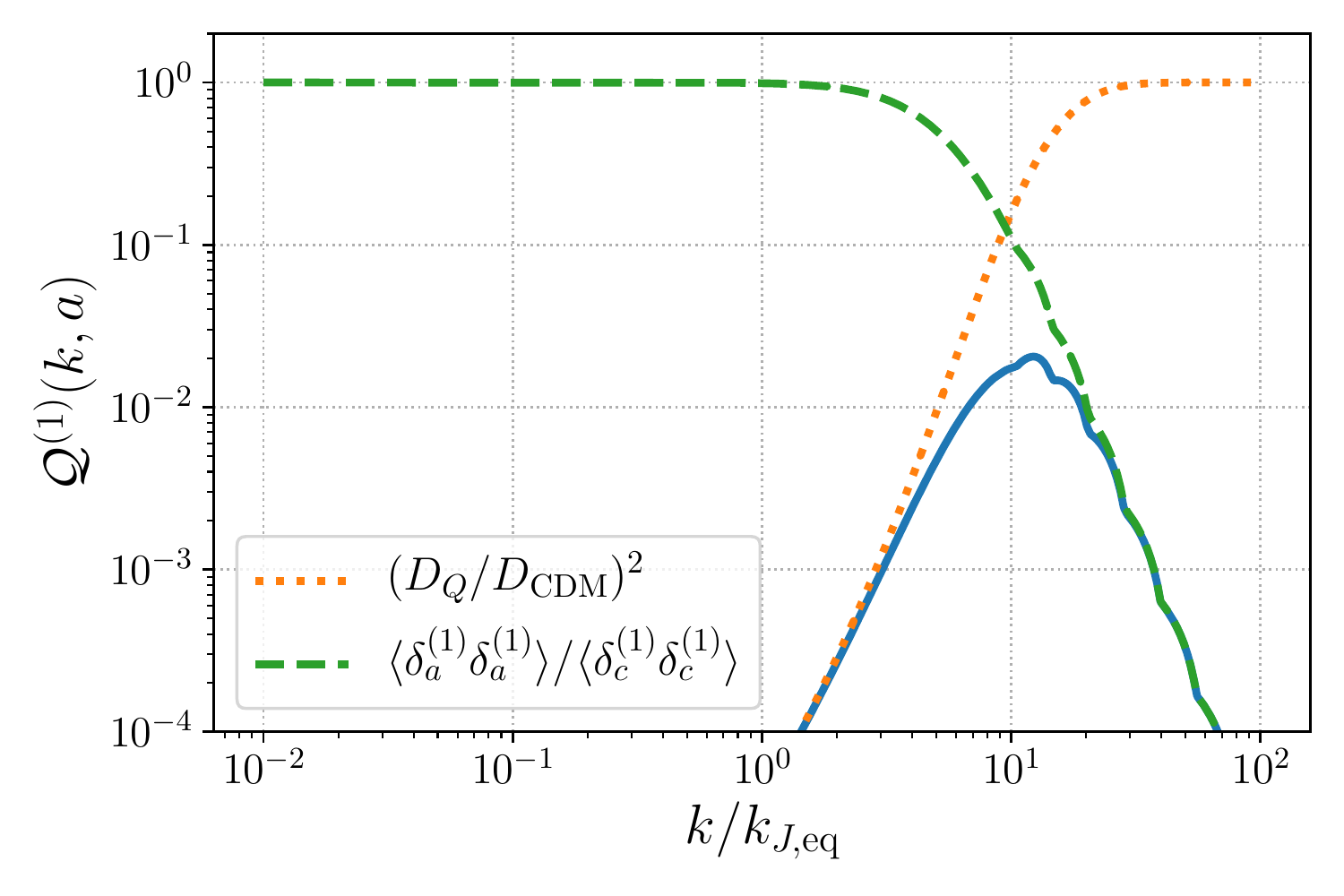}
    \caption{Amplitude of the wave effects at linear order. We see that the wave effects begin to dominate in a regime where the axion perturbations are severely suppressed and that the relative size of the quantum corrections are bounded from above. This was generated for an axion mass of $10^{-27}$ eV at redshift $z=1$.}
    \label{fig:lin_qc}
\end{figure}
We plot $\mathcal{Q}^{(1)}$ in Fig.~\ref{fig:lin_qc} where we can see that the sharp cutoff in the variance of the axion perturbations keeps the amplitude of the total quantum corrections bounded. Note that we used the fitting formula of Ref~\cite{Lague2020EvolvingUltralight} Eq.~(23) to remove numerical artifacts when computing the growth factor directly from Bessel functions.

It remains to show that this reasoning applies to higher order perturbations e.g. $\delta^{(2)},\;\delta^{(3)}$, etc. This is due to the fact that higher order perturbations are computed using lower order ones through recursion relations. Therefore if $\delta^{(1)}_a(k)= 0$ for $k\gg k_{J,\mathrm{eq}}$, then so will $\delta_a^{(2)}(k)$ and  $\delta_a^{(3)}(k)$. With regards to the growth, the second order growth factor can be found by solving the inhomogeneous equation~\cite{Bernardeau2002Large-Scale}
\begin{align}
    \ddot{D}_2(k,a) + 2H\dot{D}_2(k,a) +\left(\frac{\hbar^2k^4}{4m_a^2a^2}-4\pi G \bar{\rho}\right)D_2(k,a) = -4\pi G \bar{\rho}D(k,a)^2.
\end{align}
We anticipate $D_2$ to behave in a similar way to the linear growth on small scales ($D_2\to 0$) and on large scales ($D_2\to D_{2,\mathrm{CDM}}$). We thus expect that the quantum corrections to the 1-loop power spectrum (which are the higher order equivalent to $\mathcal{Q}^{(1)}$) to be small as well. This is corroborated by the analysis of Ref.~\cite{Li2019NumericalModel} where it was shown that the wave effects on higher order corrections become important at scales at which the 1-loop power spectrum is heavily suppressed. It is worth pointing out that this study was conducted for a single component dark matter of $m_a=10^{-23}$ eV. In the case of a two-component dark matter, the total dark matter perturbations ($\delta_d$) obey 
\begin{align}
    \left \langle \delta_d(k) \delta_d(k) \right \rangle &=  \left \langle \left[\frac{\Omega_a}{\Omega_d}\delta_a(k) + \left(1-\frac{\Omega_a}{\Omega_d}\right)\delta_c(k)\right] \left[\frac{\Omega_a}{\Omega_d}\delta_a(k) + \left(1-\frac{\Omega_a}{\Omega_d}\right)\delta_c(k)\right]\right \rangle \\&= \left(\frac{\Omega_a}{\Omega_d}\right)^2 \left\langle\delta_a(k) \delta_a(k) \right\rangle + 2\left(1-\frac{\Omega_a}{\Omega_d}\right)\left(\frac{\Omega_a}{\Omega_d}\right) \left\langle\delta_a(k) \delta_c(k) \right\rangle \nonumber \\&\;\;\;\;   + \left(1-\frac{\Omega_a}{\Omega_d}\right)^2 \left\langle\delta_c(k) \delta_c(k) \right\rangle.
\end{align}
Therefore the total one-loop contribution to the dark matter gives
\begin{align}
    P_\mathrm{1-loop}^{dd}(k) = \left(\frac{\Omega_a}{\Omega_d}\right)^2 P_\mathrm{1-loop}^{aa}(k) + 2\left(\frac{\Omega_a}{\Omega_d}\right)\left(1-\frac{\Omega_a}{\Omega_d}\right) P_\mathrm{1-loop}^{ac}(k) + \left(1-\frac{\Omega_a}{\Omega_d}\right)^2 P_\mathrm{1-loop}^{cc}(k).
\end{align}
Since only $P^{aa}$ and $P^{ac}$ have contributions from the wave effects, we see that the amplitude of these effects in the mixed case are multiplied by at least one power of $\Omega_a/\Omega_d$ which is less than 0.05 in the vast majority of the cases in our analysis. We conclude that the corrections from the wave effects on axion dynamics are even smaller in the mixed case and vanish in the limit $\Omega_a\ll 1$, as expected.

\section{Axions' Anisotropic Effects}\label{app:anisotropic}
In the rightmost panel of Fig.~\ref{fig:mon_quad_model}, we observe an enhancement of the quadrupole moment despite a suppression of structure for the $\ell=0$ multipole. This is a unique feature as it indicates possible anisotropic effects of the structure suppression of axions and constitutes a completely new signature beyond the well-known structure suppression. To investigate this, we make use of a very simple redshift space model for the galaxy power spectrum where we approximate the galaxy power spectrum as
\begin{align}
    P_g(k,\mu) \approx  e^{-(k\mu f \sigma_v)^2} \left(1+f\mu^2\right)^2 b_g^2 P_\mathrm{lin}(k),
\end{align}
where $b_g$ is the galaxy bias and where $\sigma_v$ is the galaxy velocity dispersion. This model is based on the Kaiser approximation~\cite{Kaiser1987ClusteringIn} with a Gaussian kernel for the finger-of-God effects. The velocity dispersion can be roughly approximated at linear order with (see Ref.~\cite{Hamilton1998LinearRedshift} and references therein)
\begin{align}
    \sigma_{v,\mathrm{lin}}^2 =\frac{1}{6\pi^2} \int dq P_\mathrm{lin}(q). \label{eq:sigma_v_lin}
\end{align}
Using Eq.~(\ref{eq:P_ell}), we have that the multipoles of the power spectrum are
\begin{align}
    P_\ell(k) = \frac{2\ell+1}{2}b_g^2 P_\mathrm{lin}(k) \underbrace{\int_{-1}^{1}d\mu\; e^{-(k\mu f \sigma_v)^2} \left(1+f\mu^2\right)^2 \mathcal{P}_\ell(\mu)}_{\equiv \mathcal{B}_\ell(k;\sigma_v)}. \label{eq:mathcal_B}
\end{align}
\begin{figure}
    \centering
    \includegraphics[width=0.6\linewidth]{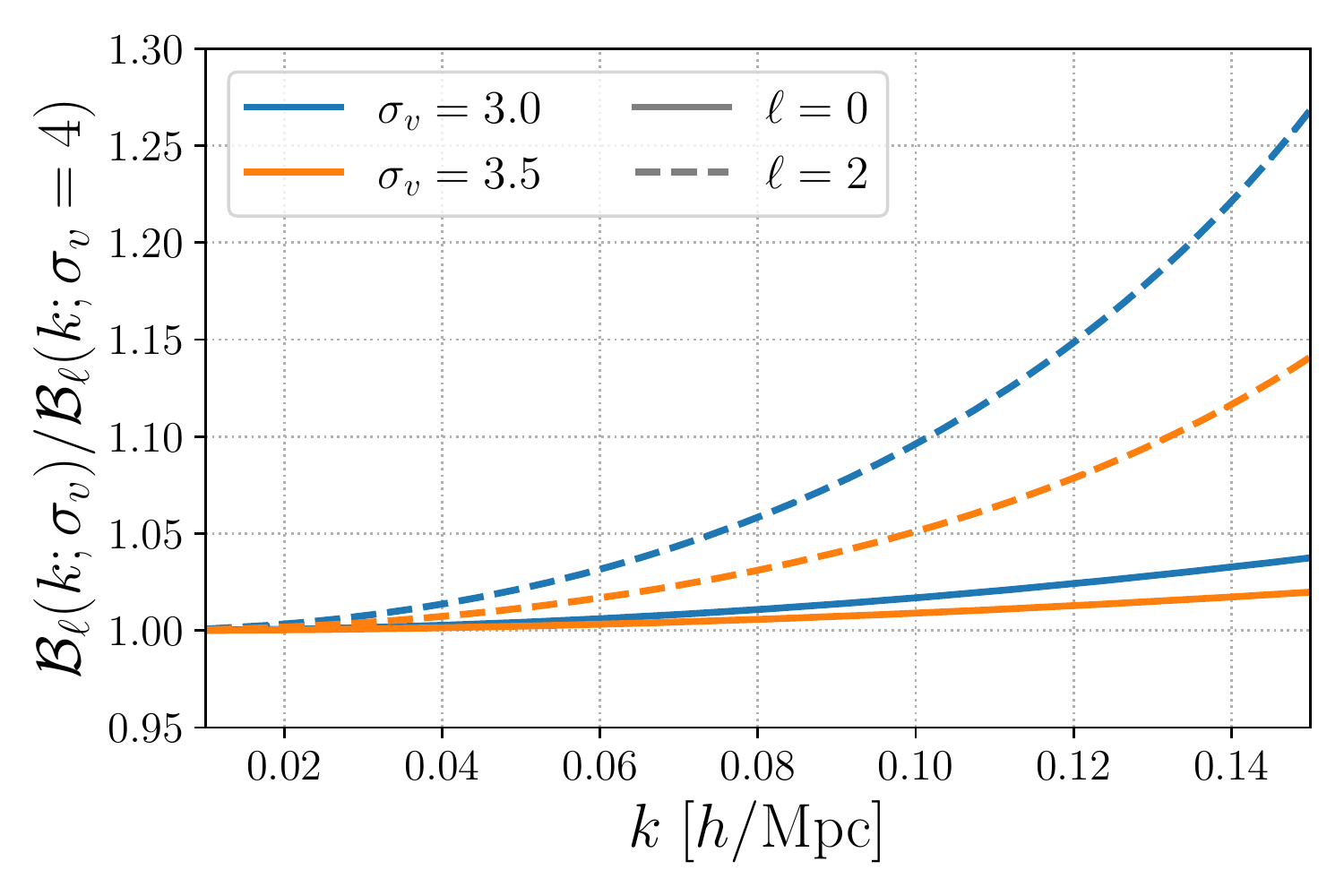}
    \caption{Function $\mathcal{B}_\ell$ as defined in Eq.~(\ref{eq:mathcal_B}) as a function of $k$ for the monopole and quadrupole. We note an enhancement on small scales when reducing the velocity dispersion (e.g. by suppressing the matter power spectrum).}
    \label{fig:B_ell}
\end{figure}

From this simple model, we find that the increase in the quadrupole moment is attributable to a decrease in the velocity divergence which arises when the linear matter power spectrum is suppressed and the value of the integral in Eq.~(\ref{eq:sigma_v_lin}) decreases. This decrease in $\sigma_v$ leads to a higher value of the $B_\ell$, especially for the $\ell=2$ as shown in Fig.~\ref{fig:B_ell}.

\section{Axion Transfer Function Interpolation}\label{app:axion_interpolation}
The axion transfer function defined in Eq.~\ref{eq:transfer} captures the deviation from $\Lambda$CDM due to axions in the matter power spectrum. It is most often obtained through semi-analytic approximations~\cite{Hu2000FuzzyCold} or numerically with adapted Boltzmann codes. In the present study however, none of those options are optimal. The approach given in Ref.~\cite{Hu2000FuzzyCold} was only for the case of $\Omega_a/\Omega_d = 1$ and the \texttt{axionCAMB} has a long runtime of up to five seconds. Although the code was used extensively for CMB studies and Fisher forecasts analyses, the higher dimensionality of the current study makes the high runtime problematic as larger number of samples is required for the MCMC chains to converge.

\begin{figure}
    \centering
    \begin{subfigure}[t]{1\textwidth}
        \centering
        \includegraphics[width=\textwidth]{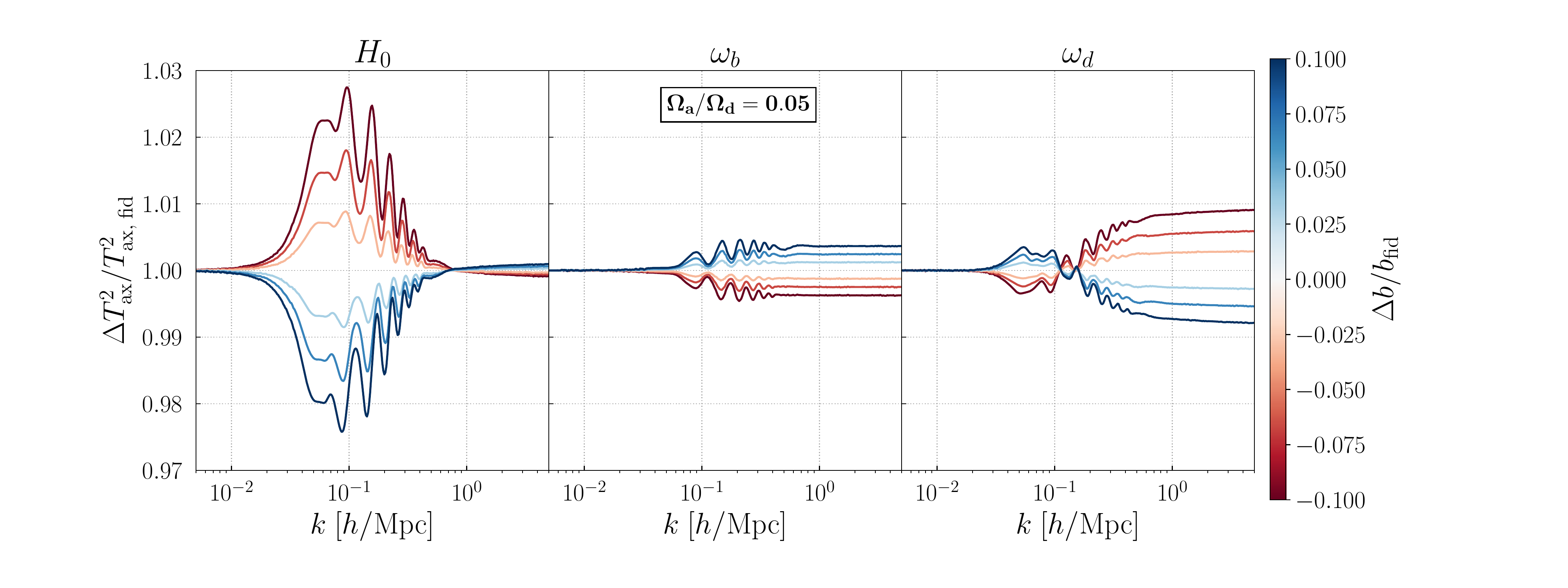}
        \caption{}
        \label{fig:T_ax_cosmo_params_005}
    \end{subfigure}
    \begin{subfigure}[b]{1\textwidth}
        \centering
        \includegraphics[width=\textwidth]{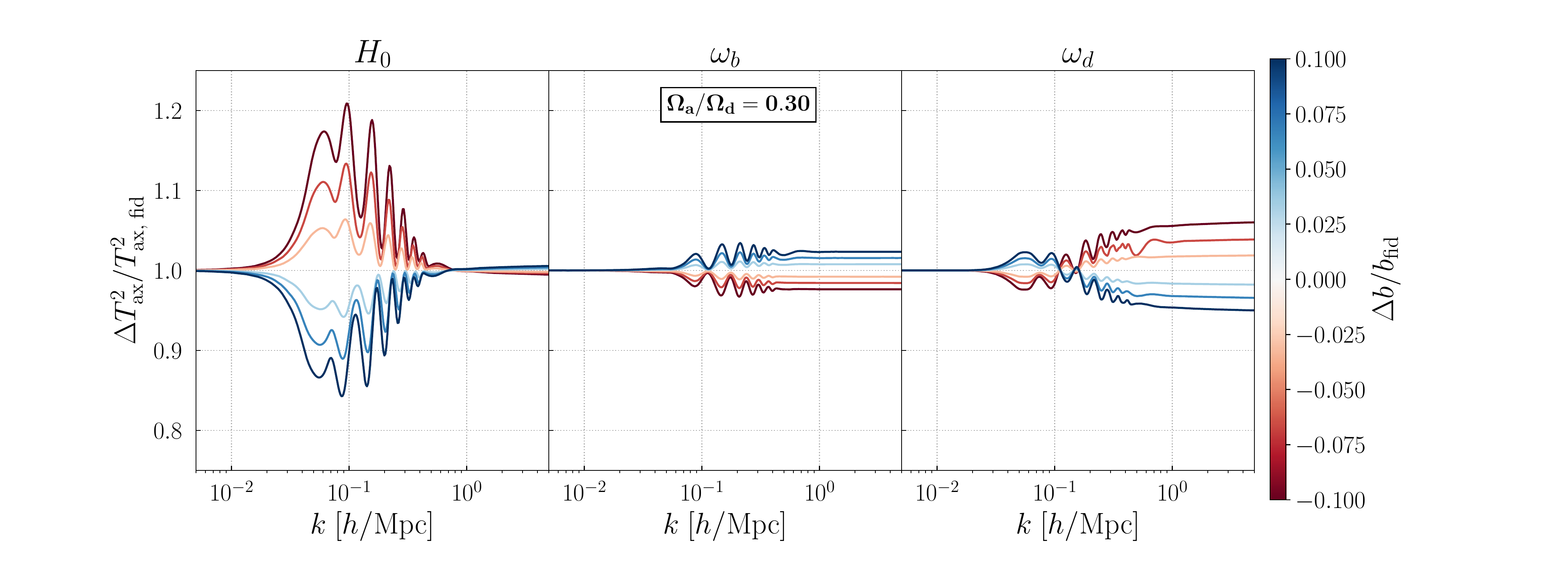}
        \caption{}
        \label{fig:T_ax_cosmo_params_030}
    \end{subfigure}
    \caption{Variation of the axion transfer function when fixing axion mass and density while perturbing cosmological parameters about their fiducial values. \textit{Top}: Variation with an axion density of 5\% with $m_a=10^{-27}$ eV. \textit{Bottom}: Variation with an axion density of 30\% at the same axion mass.}
\end{figure}

To overcome the computation time issue, we use an interpolating approach where the result of the \texttt{axionCAMB} code is calculated for a predefined grid in parameter space. The the \texttt{RegularGridInterpolator} of the \texttt{Scipy} package\footnote{\url{https://docs.scipy.org/doc/scipy/reference/generated/scipy.interpolate.RegularGridInterpolator.html}} is used to linearly interpolate the axion transfer function for the choice of parameters given by the MCMC sampler. Two approaches are considered when building the interpolator. The first is to assume the axion transfer function depends only on the axion parameters. In other words, at a fixed mass $m_a$, one only need to interpolate over different values of $\Omega_a/\Omega_d$. This approximation is valid in the limit where the axion fraction is very small since
\begin{align}
    T_\mathrm{ax}^2(\Omega_a/\Omega_d, \mathbf{b}_\mathrm{cosmo}; k) = \frac{P(\Omega_a/\Omega_d, \mathbf{b}_\mathrm{cosmo}; k)}{P(\Omega_a/\Omega_d=0, \mathbf{b}_\mathrm{cosmo}; k)} \to 1 \;\;\;\mathrm{when}\;\;\; \Omega_a/\Omega_d\to 0\;\;\; \forall \; \mathbf{b}_\mathrm{cosmo},\; k,
\end{align}
where $\mathbf{b}_\mathrm{cosmo}$ denotes the vector of cosmological parameters without including $A_s$\footnote{Although the power spectrum amplitude is varied in this study, it does not affect the transfer function which is a ratio of power spectra.}. So we expect the axion transfer function to be independent of cosmological parameters at sufficiently small concentrations. We note that the prior given for the axion fraction includes a wide range of values for which this does not necessarily hold. This observation is illustrated by the difference in transfer function variation between Fig.~\ref{fig:T_ax_cosmo_params_005} and Fig.~\ref{fig:T_ax_cosmo_params_030}.

The interpolator can be corrected by including prefactors accounting for the correlation between the axion density and the cosmological parameters.
\begin{align}
    T_\mathrm{ax,\;corrected}(\mathbf{b}_\mathrm{axion}, \mathbf{b}_\mathrm{cosmo}; k) = \prod_{b_i\in \mathbf{b}_\mathrm{cosmo} }\mathcal{C}_{b_i}(b;k)\times T_\mathrm{ax,\;naive}(\mathbf{b}_\mathrm{axion}; k),
\end{align}
where
\begin{align}
    \mathcal{C}_{b_i}(b,k) \equiv \frac{T_\mathrm{ax}(\mathbf{b}_\mathrm{axion},b_i=b;k)}{T_\mathrm{ax,\;fid}(\mathbf{b}_\mathrm{axion};k)}.
\end{align}
\begin{figure}
    \centering
    \includegraphics[width=0.6\linewidth]{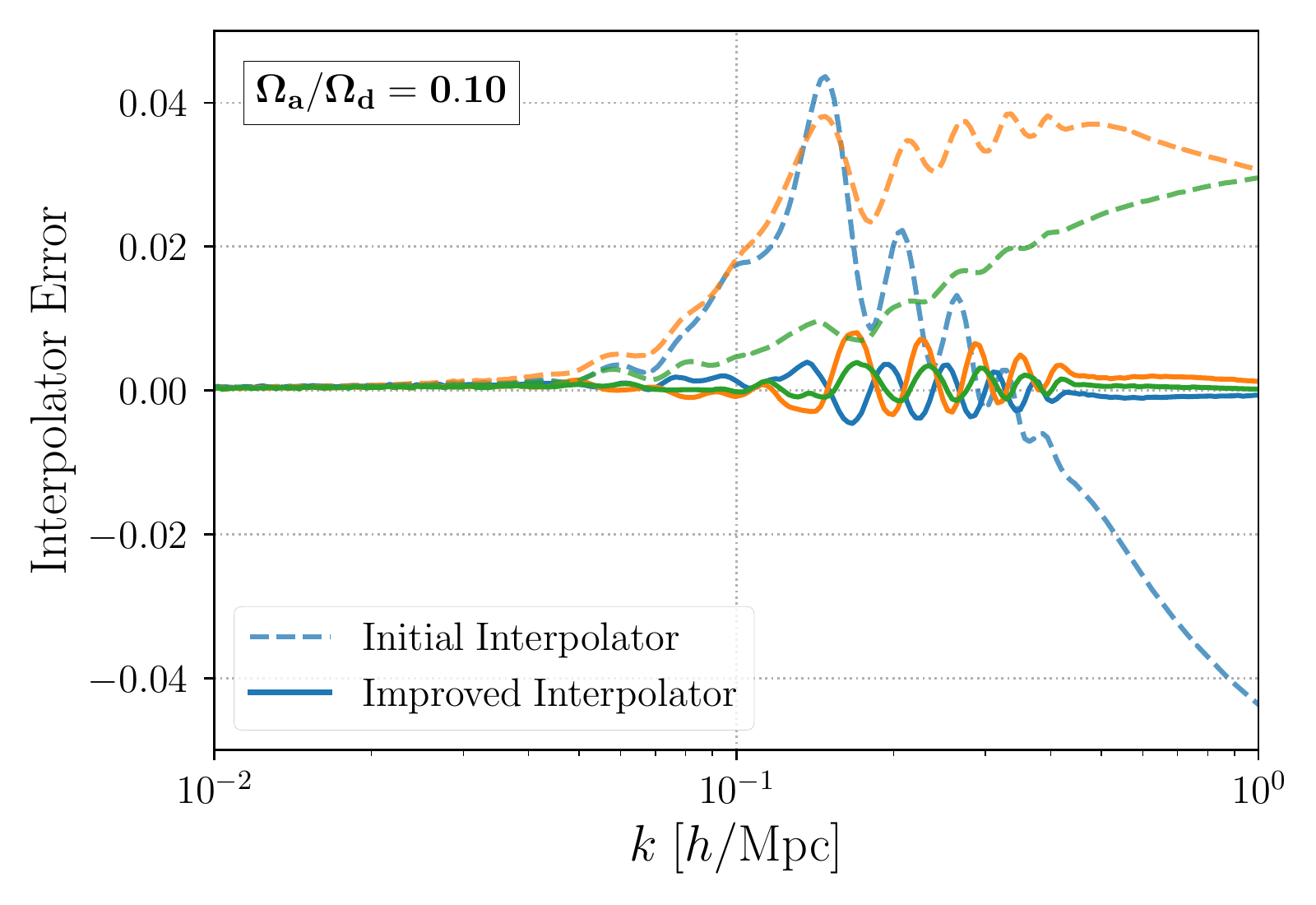}
    \caption{Relative error from the two interpolation method described in this section with respect to the \texttt{axionCAMB} result. The initial interpolation is done assuming the transfer function depends only on axion parameters and is independent of cosmological ones.}
    \label{fig:3_sample_cosmo_interp}
\end{figure}
In other words, we create four distinct interpolators. The first three vary a single cosmological parameter along with the axion fraction and evaluates the axion transfer function as a function of scale up to $k=0.5\;h/$Mpc. We then divide the result by the axion transfer function with the same axion fraction and at the fiducial values of \textit{all} the cosmological parameters. This calculation captures the effect of varying both a cosmological parameter and the axion fraction simultaneously. The normalization ensures that $\mathcal{C}_{b_i}(b,k)\to 1$ when $b\to b_\mathrm{fid}$. After calculating the correction factors for $\{\omega_b, \omega_d, H_0\}$, we calculate the naive axion transfer function which accounts for the impact of axions while keeping the cosmological parameters constant. The product of the four captures the effects of axions and the correlation between axions and cosmological parameters. Note that this process is repeated for each axion mass bin as the axion mass is fixed for each run. Accounting for these small correction improves the accuracy of the interpolation scheme as shown in Fig.~\ref{fig:3_sample_cosmo_interp} where the transfer function was calculated for three randomly sampled choices of cosmological parameters and compared to the \texttt{axionCAMB} result. This improved scheme allows the interpolation to maintain percent level error with the \texttt{axionCAMB} output at an axion concentration of $\Omega_a/\Omega_d\lesssim 0.1$ which, as shown in Fig.~\ref{fig:combined_constraint}, is where more than 95\% of the samples are taken.

The final step in using the interpolator is to multiply the resulting transfer function with the \texttt{Class} matter power spectrum which interfaces well with the \texttt{PyBird} functionalities. By not requiring the \texttt{high-precision} parameter in the input file, we get an average runtime per execution of 2.7 seconds. The combination of the \texttt{Class} and the interpolation scheme is .7 s per execution which represents an improvement by a factor of 3.85 per iteration. We run the \texttt{axionCAMB} for the interpolator for 40 different choices of each cosmological parameters and for 80 axion fractions. Combined with the naive transfer function which requires 80 executions, we arrive at a total of $\sim 10^4$ executions of the Boltzmann code for the creation of the interpolation tables. Assuming an average $\sim 2\times 10^5$ samples for the chains to converge, we get that the total runtime on a single core for the Boltzmann code approach is of $5.4\times 10^{5}$ s per chain while the interpolation scheme is at $1.7 \times 10^5$ s which represents a reduction of about 70\% in total computational cost.



\bibliographystyle{unsrt}
\bibliography{bibliography.bib}








\end{document}